\documentclass[12pt,preprint]{aastex}
\usepackage{bm}  
\usepackage{amsmath}

\newcommand{\cf}{cf.}
\newcommand{\eg}{e.g.}
\newcommand{\ie}{i.e.}
\newcommand{\etal}{et al.}
\newcommand{\bmu}{{\bm{\mu}}}
\newcommand{\bpi}{{\bm{\pi}}}
\newcommand{\bgamma}{{\bm{\gamma}}}
\newcommand{\e}{{\rm{E}}}
\newcommand{\cE}{{\mathcal{E}}}

\newcommand{\Rot}{{\mathbb{R}}}
\newcommand{\vers}[1]{{\hat{\bm{#1}}}}
\newcommand{\peri}{{\rm peri}}
\newcommand{\kep}{{\rm kep}}
\newcommand{\node}{{\rm node}}
\newcommand{\ba}{{\bf a}}
\newcommand{\bvv}{{\bf v}}
\newcommand{\br}{{\bf r}}
\newcommand{\dd}{{\rm d}}

\shorttitle{OGLE-2009-BLG-020 verifiable mass, 
distance and orbit predictions.}
\shortauthors{Skowron et al.}

\begin{document}


\title{Binary microlensing event OGLE-2009-BLG-020 gives a 
verifiable mass, distance and orbit predictions.}


\author{
J.~Skowron\altaffilmark{1},           
A.~Udalski\altaffilmark{4,51},        
A.~Gould\altaffilmark{1,50},          
Subo~Dong\altaffilmark{2,3,50},       
L.~A.~G.~Monard\altaffilmark{7,50},   
C.~Han\altaffilmark{14,50},           
C.~R.~Nelson\altaffilmark{1,50},      
J.~McCormick\altaffilmark{9,50},      
D.~Moorhouse\altaffilmark{8,50},      
G.~Thornley\altaffilmark{8,50}
% ...
\\
}
\email{jskowron@astronomy.ohio-state.edu}



\altaffiltext{1}{Department of Astronomy, Ohio State University, 
                 140 W. 18th Ave., Columbus, OH 43210, USA}
\altaffiltext{2}{Institute for Advanced Study, 
                 Einstein Drive, 
                 Princeton, NJ 08540, USA}
\altaffiltext{3}{Sagan Fellow}
\altaffiltext{4}{Warsaw University Observatory, Al. Ujazdowskie 4, 
                 00-478 Warszawa, Poland}
% ...
\altaffiltext{7}{Bronberg Observatory, 
                 Centre for Backyard Astrophysics Pretoria, 
                 South Africa}
\altaffiltext{8}{Kumeu Observatory, Kumeu, New Zealand}
\altaffiltext{9}{Farm Cove Observatory, 
                 Centre for Backyard Astrophysics, 
                 Pakuranga, Auckland, New Zealand}
% ...
\altaffiltext{14}{Department of Physics, Chungbuk National University, 
                  Cheongju 361-763, Republic of Korea}
% ...
\altaffiltext{50}{Microlensing Follow Up Network ($\mu$FUN)}
\altaffiltext{51}{Optical Gravitational Lens Experiment (OGLE)}
\altaffiltext{52}{Microlensing Observations in Astrophysics (MOA) Collaboration}
% ...


\begin{abstract}
We present the first example of binary microlensing for
which the parameter measurements can be verified
(or contradicted) by future Doppler observations.
This test is made possible by a confluence of two
relatively unusual circumstances.  First, the binary
lens is bright enough ($I=15.6$) to permit Doppler
measurements. Second, we measure not only the usual
7 binary-lens parameters, but also the ``microlens parallax''
(which yields the binary mass) and two
components of the instantaneous orbital velocity.
Thus we measure, effectively, 6 'Kepler+1' parameters
(two instantaneous positions, two instantaneous velocities,
the binary total mass, and the mass ratio).
Since Doppler observations of the brighter binary component determine
5 Kepler parameters (period, velocity amplitude, eccentricity,
phase, and position of periapsis), while the same
spectroscopy yields the mass of the primary, the combined
Doppler + microlensing observations would be overconstrained
by $6 + (5 + 1) - (7 + 1) = 4$ degrees of freedom.
This makes possible an extremely strong test of the microlensing
solution.

We also introduce a uniform microlensing notation 
for single and binary lenses, we define conventions, summarize all 
known microlensing degeneracies and extend a set of parameters 
to describe full Keplerian motion of the binary lenses. 
\end{abstract}


\keywords{Galaxy: bulge -- gravitational lensing -- stars: binary}

\section{Introduction}
\label{s:intro}

Gravitational microlensing is nowadays a well established method for 
discovering binary and planetary systems \cite[\eg][]{gould09,gaudi10}. By 
analyzing flux variations in time, microlensing measures
a wide range of system parameters including the distance and mass of 
the binary, its orbital and proper motion as well as the mass ratio 
and separation of its components \cite[\cf][]{dong09a, bennett10}. 

Previous studies of gravitational microlensing events 
have had little or no possibility for {\it post factum} observational 
confirmation of derived system parameters, since usually the stars 
involved in this one-time event are too faint and too distant to be  
within reach of current astrometric or spectroscopic instruments.
Some exceptions come from astrometric confirmation of the nature of 
a single lens event with direct imaging done with Hubble Space 
Telescope (HST) \citep{alcock01, gould04b, kozlowski07}. 
Also in some cases it was possible to take spectra of the 
microlensed source that confirm the microlensing interpretation 
\citep{gaudi08b}. But the fact of an event being caused by 
microlensing is the main subject of these tests rather than the 
values of previously derived parameters.
This does not mean that the microlensing measurements have no tests. 
Indeed there are many self-consistency checks including the 
agreement of mass and distance with the amount of light coming to 
us \cite[\eg][]{gaudi08a,bennett10}, as well as testing of the derived
values of the orbital parameters being consistent with bounded Keplerian 
orbits (\eg\ \citealt{sumi10}, Batista \etal\ 2011 in prep.,
Yee \etal\ 2011 in prep.). 
The only issue is the shortage of possibilities to verify the 
results with  independent or direct observations.

In this work we present analysis of one microlensing event 
(OGLE-2009-BLG-020\footnote{\url{http://ogle.astrouw.edu.pl/ogle3/ews/2009/blg-020.html}})
caused by a $\sim 1.1 \, M_\Sun$ binary system 
passing near the line of sight to an ordinary red giant star. 
We derive system parameters using the same, standard methods that 
are applied to other binary and planetary microlensing events. 
Because of a peculiarity of this event, \ie, that it is caused by 
a relatively close by ($\approx 1.1$ kpc) and bright ($I=15.6$) 
binary star, there is a possibility of direct verification 
of the derived parameters with follow-up spectroscopic measurements.

The data gathered on this event are described in Section \ref{s:data}.
The fitting of the microlensing model to the light curve is 
presented in Section \ref{s:modelling}, and the physical parameters 
of the system are calculated in Section \ref{s:params}. 
In Section \ref{s:discussion} we 
discuss the results and, in particular, in Section \S\ref{s:test} we 
present how the microlensing solutions
can be tested by radial velocity measurements.
In the Conclusions (\S\ref{s:conclusions}) we advocate for radial 
velocity follow-up observations to confirm the nature of the event 
and its parameters, which constitute a general test of the 
microlensing method, in particular the accuracy of the parameters 
currently being derived for microlensing planets.

In Appendix \ref{a:notation} we present
microlensing parameters together with all required conventions, 
introduce a uniform microlensing notation,
extend work of \citet{gould00} by introducing parameters
describing full Keplerian motion of the lens components, 
and review symmetries known in microlensing.
In Appendix~\ref{a:jacobian} we derive the transformation between the
microlensing and Keplerian orbit parameters.

\section{Observational data}
\label{s:data}

\subsection{Collection}
On 15 February 2009, Heliocentric Julian Date (HJD) $\sim$ 
2454878, the Optical Gravitational Lensing Experiment (OGLE) team 
announced ongoing microlensing event OGLE-2009-BLG-020 detected 
by the Early Warning System 
(EWS)\footnote{\url{http://ogle.astrouw.edu.pl/ogle3/ews/ews.html}} 
and observed on the 1.3m Warsaw Telescope in Las Campanas Observatory in 
Chile. The event was also monitored by the Microlensing Observations
in Astrophysics (MOA) 1.8m telescope at Mt. John University
Observatory in New Zealand. 
On HJD$'$ $\sim$ 4915 
($HJD' = HJD - 2450000$) it could be seen that the light curve was 
deviating from the standard \citet{paczynski} model, and follow-up 
observations by other telescopes began: First with the 2.0m 
Faulkes South (FTS) telescope in Siding Spring, Australia and the 2.0m 
Faulkes North (FTN) telescope in Haleakala, Hawaii operated by 
RoboNet\footnote{\url{http://robonet.lcogt.net/}}; the 36cm telescope 
at Kumeu Observatory\footnote{\url{http://kumeu.blogspot.com/}}, 
New Zealand and the 36cm telescope at Bronberg Observatory, Pretoria, 
South Africa as a part of Microlensing Follow-Up Network 
($\mu$FUN)\footnote{\url{http://www.astronomy.ohio-state.edu/microfun/}}. 
Then, shortly before the first caustic crossing occurred 
(HJD$'$ $\sim$ 4917.3) observations began on the 1m telescope 
on Mt. Canopus belonging to University of Tasmania (part of PLANET 
Collaboration) and the 36cm telescope in Farm Cove 
Observatory\footnote{\url{http://www.farmcoveobs.co.nz/}} 
(also $\mu$FUN). Essential data near highest magnification 
(HJD$'$ $\sim$ 4917.6, March 27th) were gathered by the Bronberg 
telescope. During the caustic exit (HJD$'$ $\sim$ 4917.75) there were 
observations started on the SMARTS 1.3m Cerro Tololo Inter-American Observatory 
(CTIO) and the 40cm telescope in Campo Catino Austral Observatory 
(CAO)\footnote{\url{http://www.campocatinobservatory.org/}}, Chile. 
Some data where gathered by other observers but too short coverage 
of the event or, in some cases, big error bars
prevented us from including these in this analysis.
Intense monitoring of the event was performed until HJD$'$ $\sim$ 4920. 

The whole light curve consists of 9 years of data with 121 days 
during the course of the visible magnification, of which 5 days 
constitute intense follow-up observations. The OGLE telescope
performed observations in $V$ and $I$ bands,
and the CTIO telescope in $V$, $I$ and $H$ bands, which permit
measurements of the color of the magnified source star. 
Together we have 5333 data points in the light curve with 2247 
during the course of the magnification event.

We take the OGLE-III V- and I-band light curves from the
project’s final data reductions \citep{udalski08}
and calibrate against the Galactic bulge photometric maps 
(Szyma{\'n}ski \etal\ 2011 in prep.), so that the OGLE magnitudes reported 
in this paper are standard V (Johnson) and I (Cousins) magnitudes.

The Bronberg data were reduced using MicroFUN's image
subtraction pipelines based on \citet{wozniak00}.
The MOA data were reduced using the survey image
subtraction pipeline \citep{bond01}.
CTIO, CAO, FCO and Kumeu data were reduced using DoPHOT package 
\cite{schechter93}. The UTAS, FTN and FTS data were reduced
using PySIS \citep{albrow09}.

We remove the effects of a differential extinction from the unfiltered
Bronberg data by using light curves of non-variable field stars
that have similar color to that of the source
\citep[for details see][]{dong09a}.

The CTIO telescope's camera, ANDICAM, takes simultaneous exposures 
in $H$- and $I$-band. Because the data quality in $H$-band is lower 
than that in $I$-band, we include only the
latter into the microlensing fit. However, we use the $H$-band data to
calculate $(I-H)$ color of
the source. To do so we cross-match the $H$-band data with the 2MASS 
catalog\footnote{\url{http://www.ipac.caltech.edu/2mass/overview/access.html}}
and calibrate neighboring stars of the lens/source to 2MASS $H$-band 
magnitudes.

\subsection{Preparation}

As calculation of microlensing magnification is very time consuming,
we bin the data to speed up the fitting procedure. On every part of
the light curve and for every observatory individually we carefully 
choose the number of bins and their start and end points depending on 
the local slope of the light curve and specifically avoid incorporating
seasonal gap (for wide bins) and daily gaps (for short bins) into 
any bin. 
Before the actual binning we fit straight lines to the points in 
the planned bins and check that the points look consistent with the lines.
We also check that the slopes for adjacent bins are comparable.
We discard 3-$\sigma$ outliers from the straight line fits as
well as all data points with observational error-bar bigger than
5 times median of the error-bars of nearby points.
Our human supervised binning procedure coupled with extensive use of 
helper algorithms yields 374 binned data points in total.

We rescale the error bars to have $\chi_d^2/dof \sim 1$ for each
separate data set (usually coming from different observatories). 
This is an iterative process wherein we use our 
best-fit model to evaluate $\chi_{d,i}^2$ values for each of the 
data points (where $d$ enumerates data sets and $i$ is the 
index of the given data point in this set), then based on these
values we find coefficients $Y_d$ and $S_d$ (as described below) 
needed to rescale error bars with the formula: 
$\sigma_{d,i}^{new} = Y_d \sqrt{\sigma_{d,i}^2 + S_d^2}$,
where $\sigma_{d,i}$ and $\sigma_{d,i}^{new}$ are the uncertainties
before and after rescaling.
After rescaling we repeat the fitting procedure and find the new values 
of the coefficients.

To find values of the rescaling coefficients, for each data set $d$,
we sort all points by the magnification given by the microlensing model. 
Then we construct 
the cumulative distribution of $\chi_{d,i}^2$ as a function of
the sorted index.
The goal is to choose the $S_d$ to make the cumulative distribution 
of $\chi_{d,i}^2$ a straight line. Then we can choose the scale 
$Y_d$ to match the requirement of $\chi_d^2/dof \sim 1$.

If we see some systematic variations in the residuals
from the best-fit model that are unlikely to be caused by the 
microlensing phenomena, we investigate whether the deviations are 
supported by more than one dataset or whether it is likely that 
observations were affected by high airmass or large seeing. 
If so, we remove these data as outliers.

\begin{figure}
\plotone{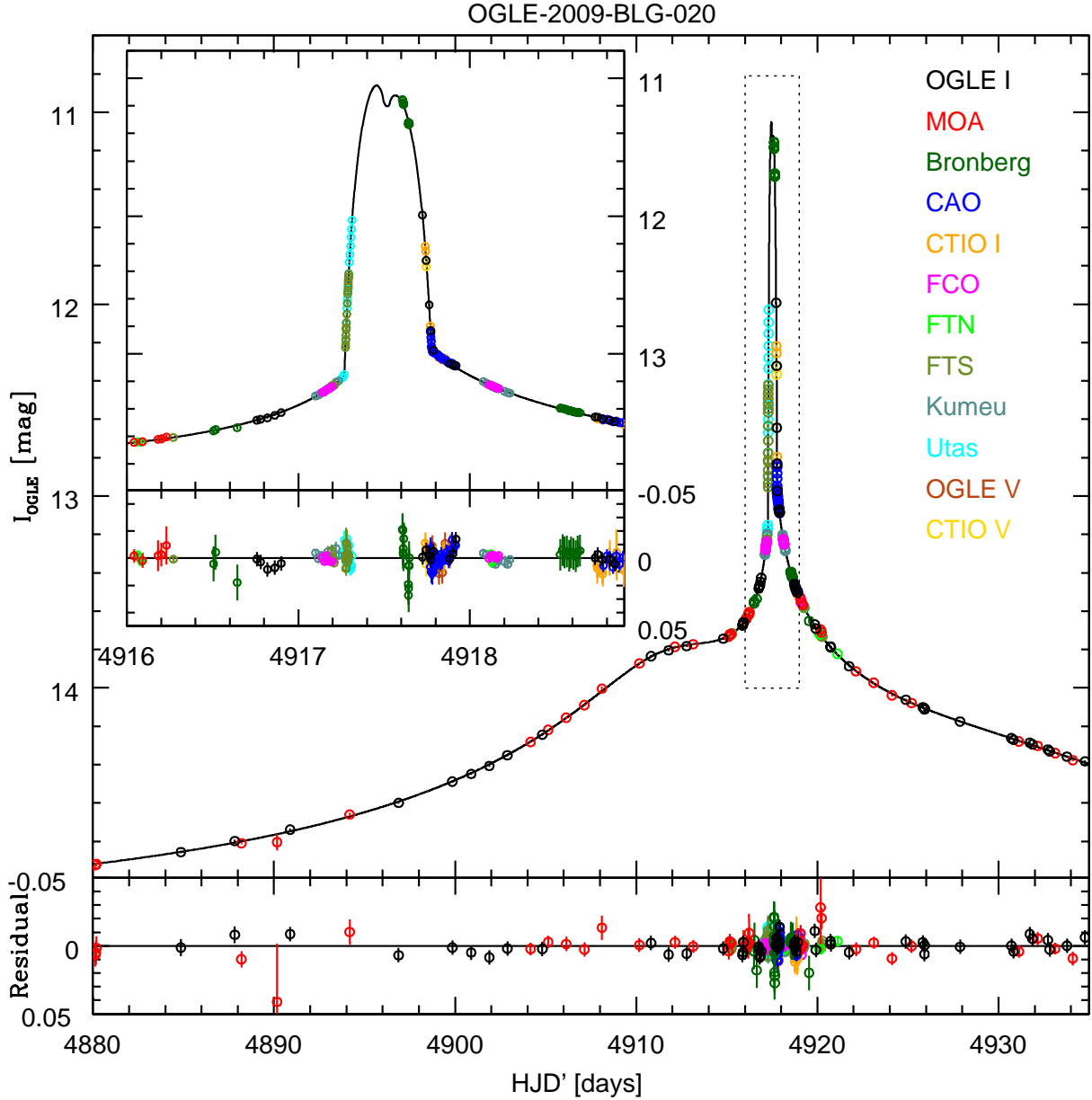}
\caption{Light curve of OGLE-2009-BLG-020. 
Different colors denote 12 data sets and the solid line shows the
best microlensing model.
The inset presents the part of the light curve featuring both 
caustic crossings.
\label{f:lc-all}}
\end{figure}

Figure \ref{f:lc-all} presents the light curve of the event.
All data points are aligned to the best fit microlensing model.
Inset shows the portion of the light curve
during highest magnification -- when the source is crossing the caustics.



\section{Microlensing model}
\label{s:modelling}

To explain the observed variation in brightness, we fit a binary 
microlensing model to the light curve. In this scenario the light 
from a distant source is bent by the binary star (lens) crossing 
near the light of sight, causing apparent brightening of the source.

In this work we closely follow the notation presented by \citet{gould00},
where $D_l$ denotes the distance to the lens, $D_s$ the distance to the source, 
$M$ the total mass of the lens, $r_\e$ the Einstein radius, 
and $\theta_\e$ the angular Einstein 
radius. 
In Appendix \ref{a:notation} we propose an extension of this notation 
to describe full orbital motion of the binary lens.

\subsection{Parametrization}
The initial mathematical model used to describe this event is 
constructed using 7 parameters: the mass ratio of the 
lensing binary ($q$), the projected separation of its components 
($s_0$) in units of the Einstein radius, the angle of the lens-source
relative motion projected onto the sky plane
with respect to the binary axis ($\alpha_0$), the Einstein crossing time 
($t_\e$), \ie, the time required for the lens to travel a distance of 
one Einstein radius, the time of the closest approach of 
the adopted center of the lens to the source ($t_0$), 
the lens-source separation at this time 
in units of the Einstein radius ($u_0$), and the source 
radius in the same units ($\rho$). See Appendix \ref{a:notation}
for conventions and full definition of all parameters.
We define the primary/secondary as the heavier/lighter component respectively.

In addition, there are two parameters 
for each observatory that connect the microlensing magnification with the 
instrumental fluxes in a given band. These are the values of the source 
flux and an additional flux not being magnified but observed in 
the same direction in the sky (a blend). If there is any light coming 
from the lens, it will be included in this blend.

To find a microlensing model we use the method developed by 
S. Dong and described in \cite{dong06} and its modification 
in \citet[Section 3]{dong09b}.

We introduce a few modifications. For example, we 
calculate the ``caustic width'' ($w$) used for the $(w,q)$ grid of 
lens geometries when searching for initial solutions. Instead of 
the planetary regime ($q \ll 1$) we are now in a stellar binary 
regime, \ie, $q \sim 1$.
We can calculate the shorter diameter of the central 
caustic with the equation:
\begin{equation}
w = \frac{4 q s^2}{ (1+q)^2 }
\end{equation}
based on work by \cite{an05}.
Unlike the formula of \citet[eq. 12]{chung05} used by \cite{dong09b}
the above formula is useful for any mass ratio, although it is accurate
only to about $20\%$ for lens separations $0<s<0.5$ and accurate to a
factor of $2$ for $s<0.7$ and stellar mass ratios 
(above $s \sim 0.7$ we are close to intermediate 
causitic geometry with resonant caustic).
Usually a much better accuracy is not required as one uses
the caustic dimensions as a basic scale for further searches, 
although for our event's geometry and chosen parametrization a more 
accurate formula is beneficial. 

We find that combining the ideas used in both formulae
gives us much more accurate results. Namely we divide the Equation (12) 
of \citet{chung05} by the factor $(1+q)^2$ making it usable 
for any mass ratio. This modified recipe is good to about $0.5\%$
for binary separations $0<s<0.5$ and to about $25\%$ for $0.5<s<0.7$,
as we get closer to resonant caustic regime.
The final formula by which we calculate the shorter diameter of the 
central caustic is derived to be:
\begin{equation}
w = \frac{4 q |\sin^3 \phi_c| }{ (1+q)^2 (s+s^{-1} - 2 \cos\phi_c)^2 }
\label{eq:w}
\end{equation}
with the parameter $\phi_c$ given by Eq. (10) in \citet{chung05}.

For wide lenses (\eg\ $s>2$ for stellar mass ratios) Equation (\ref{eq:w}) could be 
multiplied by $\sqrt{1+q}$ to produce an estimate of the short diameter 
of the caustic that is usable for our purposes but is a
less accurate approximation.

The $(1+q)^2$ factor can be used as well to modify the Equation (11) of
\citet{chung05} formula for calculating the {\it longer} diameter 
of the central caustic. Although it is usable for 
stellar binary mass ratios, it is accurate only to the factor of $2$ 
for $0<s<0.5$ ($5$ for $0.5<s<0.7$).

In our method we also modify the parametrization of the Markov Chain 
Monte Carlo (MCMC) $\chi^2$ minimization by using $(u_0/w)$ rather
than $u_0$ (impact parameter). When we observe a near-cusp caustic
crossing or cusp approach in the light curve we can expect 
that the value of $(u_0/w)$
will correlate less with the changes in lens geometry given by
$q$ and $s$. This is the case in our model for which the 
dependence of $u_0$ on $q$ and $s$ is strong but nearly vanishes 
when $u_0$ is divided by $w$. 

\subsection{Searching for solutions}
\label{s:search}

First we fit a 7 parameter microlensing model by 
\citep[after][]{dong09b} creating a broad grid in 3 parameters 
($\log q$, $\log w$, $\alpha$) and performing minimization in the 
remaining 4 parameters, which are chosen to be:
$t_0$, $(u_0/w)$, $t_{\rm eff} (\equiv u_0 t_\e)$, and $t_* (\equiv \rho t_\e)$.
The values of the fluxes (2 for each 
observatory) are calculated analytically with the least-squares method. 
With this standard parametrization we are not able to find a
satisfactory solution, in a sense where all data sets can be aligned
in a coherent way.

We then extend our model by taking into account the Earth 
orbital motion during the course of the event. This so called  
parallax effect is described with 2 additional parameters.
We use the geocentric parallax formalism \citep{an02, gould04a}, wherein we 
utilize $\bm{\pi}_\e=(\pi_{\e,N},\, \pi_{\e,E})$ as model 
parameters
(see Appendix \ref{s:pointlenshigh} for definition).
The parameters $\alpha_0$, $t_0$, and $u_0$, which previously 
described a straight trajectory of the lens in front of the source, now 
describe the trajectory we would see if the Earth's velocity was 
constant throughout the course of the event and not subject 
to gravitational acceleration. For the value of this constant
velocity we take the
real Earth velocity at some fiducial time 
$t_{0,{\rm par}} \equiv 4917.52$, and since 
the shift between real source position and the straight trajectory 
is by definition equal to zero at this chosen time, it is convenient 
to fix it close to some important features of the light curve, 
in our case it is a time between the caustic crossings.

By performing minimization in a similar manner as previously, 
but with a 9 parameter model, we find a solution capable of explaining 
the shape of the observed variability. The parameters of this model 
are gathered in Table \ref{t:params}.
We have tested whether there are other solutions emerging from known 
degeneracies. Our best model has a binary separation smaller than 
the Einstein radius. We find no solution with a ``wide'' geometry 
given by the $s \leftrightarrow s^{-1}$ degeneracy.

The magnitude of the parallax ($\pi_\e$) together with the 
angular size of the Einstein radius ($\theta_\e$) yield the total mass of the 
lens ($M$) and a distance to the lens ($D_l$) \citep[\eg][]{an02}. 
After \citet{gould00} we have:
\begin{gather}
M_l  =  \frac{\theta_\e}{\kappa\pi_\e}, 
\quad \kappa \equiv 4 G/(c^2 AU) \approx 8.1 \, \rm{mas}\,{M_\Sun}^{-1} \\
D_l = AU/\pi_l, \quad  \pi_l = \theta_\e \pi_\e + \pi_s
\end{gather}
where we derive $\theta_\e$ from the
source radius ($\rho$) in units of $r_\e$ and the measurement of 
the angular radius of the source in physical units 
($\theta_*$, see section \ref{sec:theta_star})
with the equation: $\theta_\e=\theta_*/\rho$. We assume the source to be in the
Galactic Bulge at $D_s=8$ kpc, and thus the source parallax is 
$\pi_s=0.125$ mas.

The total mass derived from the best `parallax-only' solution is $0.84 \, M_\Sun$ 
and the distance is $0.61$ kpc. This is in rough agreement with the position
of the lens (assuming that all blended light is comming from the lens) 
on the color-magnitude diagram (CMD) and the assumption 
that the binary companions belong to the main sequence.

However, there are still some structures in the residuals of the fit,
which we hope will be reduced by extending our model by
introducing new parameters describing the orbital motion of the lens.

\subsection{Expanding the model}

\subsubsection{Linear orbital motion approximation}
The canonical way of introducing orbital motion of the lens into the 
microlensing model is to add two parameters describing instantaneous
velocities in the plane of the sky of the secondary binary component
relative to the primary. As we expect that the duration of the binary 
microlensing perturbation to be only a fraction 
of the orbital period, the assumption is
that these velocities do not change much during the course of the event. 
Thus, one keeps them constant and uses a linear approximation of the 
position of the lens components as a function of time: 
$\alpha(t) = \alpha_0 -  \gamma_\perp(t-t_{0,par})$ and 
$s(t) = s_0 + \dot{s} (t-t_{0,par})$ $= s_0 (1 + \gamma_\parallel (t-t_{0,par}))$. 
Examples of microlensing 
events modeled using this approximation can be found in: 
\citet{albrow00, jaroszynski05, dong09a, ryu10, hwang10}. 
More discussion of these
parameters can be found in \S \ref{a:orbit2d}.

We fit our light curve with this, now 11 parameter, model
and notice significant improvements in the goodness of fit
($\chi^2$ changed from 370.29 to 352.70).
This indicates that orbital motion is an important effect in this
event. The fit also reveals a very strong degeneracy between 
two model parameters,
namely $\pi_{\e,N}$ and lens angular velocity ($\gamma_\perp$).
We discuss this further in Section \ref{s:piEN_omega}.

\subsubsection{Full Keplerian orbit}
Although this linear approximation works well for a large subset 
of microlensing events, in order to allow 
comparison with the radial velocity (RV) measurements it is 
profitable to use the full Keplerian orbit parametrization. 
In addition to being more accurate, the additional 
advantage of this approach is to avoid all unbound orbital
solutions (with eccentricity $\ge 1$) and to enable the introduction of priors 
on the values of orbital parameters directly into MCMC calculations,
if one decides to adopt that approach.

We describe the orbit of the secondary component relative to 
the primary by giving 3 cartesian positions and 3 velocities at one
arbitrarily chosen time ($t_{0,\kep}$). 
For convenience we utilize the same time ($t_{0,{\rm par}}$) 
used for calculating the parallax shifts ($t_{0,\kep}=t_{0,{\rm par}} = 4917.52$).
As described above (\S\ref{s:search}), our extended microlensing model 
(with parallax parameters) carries information about the mass ratio, 
the total mass of the lens, and the physical scale in the lens plane, 
so together with the six instantaneous phase-space coordinates
it comprises a complete set of system parameters (except systemic
radial velocity).
This, for example, allows us to calculate the relative RV
at any given time.

Let $r_x \equiv s_0 \cdot D_l \theta_\e$ be the projected binary 
separation in physical units.
We define the instantaneous orbital velocities after 
\citet[see Appendix]{dong09a} as $r_x \bm{\gamma}$ where
$r_x \gamma_\parallel$ and $r_x \gamma_\perp$ are the
instantaneous velocities
in the plane of the sky, parallel and perpendicular to the
projected binary axis respectively, and $r_x \gamma_z$ is 
the relative RV, \ie, the relative velocity of the two components
perpendicular to the plane of the sky. 
We note that $\gamma_\perp$ is the instantaneous angular 
velocity in the plane of the sky and $\gamma_\parallel$ is equal to
$\dot{s}/s_0$. We use the convention that $\gamma_z>0$ for movement 
toward the observer; this is opposite to convention usually 
used in RV measurements (see discussion in \S\ref{a:kep}).

Without loss of generality we set the cartesian 
coordinate system to have its first two axes in the plane of the sky,
the origin in the primary component of the lens,
and to be rotated in a way that the first axis coincides with the binary axis 
at the chosen time $t_{0,\kep}$. Then, the
3-dimensional position of the secondary at $t_{0,\kep}$ can be 
described by the vector $(s_0,0,s_z)$ and the 3-dimensional velocities 
described by $(\gamma_\parallel, \gamma_\perp, \gamma_z)$, where we 
have introduced one more parameter: the position along the line 
of sight ($s_z$) measured in units of $r_\e$. See \S \ref{a:kep}
for conventions.

In all, there are now 13 microlensing parameters, in addition to the
$2 n_{obs}$ flux parameters, where $n_{obs}=12$ is the number of 
observatories.
As this is a full description of the system, it is possible to calculate
all properties in physical units as well as
standard Kepler parameters of the orbit -- \ie, eccentricity ($e$), 
time of periapsis ($t_{peri}$), semi-major axis ($a$) and 
3 Euler angles: longitude of ascending node ($\Omega_{node}$),
inclination ($i$), and argument of periapsis ($\omega_{peri}$).
Having those we can find the exact position of the lens components at 
any given time, which in turn specifies the lens geometry,
and the projected position of the source relative to this
geometry. 

\subsection{Priors and their transformation}
\label{s:priors}
\subsubsection{Priors on orbital motion}
\label{s:orbital}
The output of the Monte Carlo Markov Chain (MCMC) consists of a set 
of points in the parameter space. We call this set ``a chain'', and
every individual solution ``a link'' in the chain. The density of 
the points in a given space bin is proportional to a likelihood
of this bin. This would be true if we assumed uniform priors on all our
fit parameters. However, if we perform a MCMC in the set of parameters 
$(x_1, x_2, x_3, ...)$, which are given by some transformation function 
from another set of parameters $(p_1, p_2, p_3, ...)$ for which we
can find physically justified priors, then those priors have to be
converted to the fit coordinates by multiplying them by a Jacobian
of this transformation: 
$||\partial(p_1, p_2, p_3, ...)/\partial(x_1, x_2, x_3, ...)||$.

We perform microlensing calculations using the 6 instantaneous 
cartesian phase-space coordinates. However, our intuition about 
priors is better in the space of Keplerian parameters. 
Thus we need to construct transformation
function and evaluate its Jacobian. In the Appendix \ref{a:jacobian}
we derive formulae for 
$j_\kep=||\partial(e, a, t_{peri}, \Omega_{node}, i, \omega_{peri})/
\partial(s_0, \alpha_0, s_z, \gamma_\parallel, \gamma_\perp, \gamma_z)||$.

We assume flat priors on values of eccentricity, 
time of periapsis, $\log a$ and $\omega_{peri}$.
To incorporate the fact that orbital angular momentum vector can be 
oriented randomly in the space, we multiply prior by 
$||\partial(\Omega_{node},\cos i)/\partial(\Omega_{node},i)||=|\sin i|$.
So our orbital motion prior, in coordinates of fit parameters, is equal to 
$j_\kep |\sin i| a^{-1}$.

\subsubsection{Priors on lens parameters}
\label{s:priors_lens}
We also include priors on the lens parameters as expected from the
simple Galactic model. 
Our trial fits to the light curve (with only Keplerian priors) show 
that the lens has substantial proper motion 
(of order $11-16\, {\rm mas}\, {\rm yr}^{-1}$) and is located  
close to the Earth ($0.5-2.4\,{\rm kpc}$). These
suggest that most of the observed proper motion is due to lens 
velocity itself (Earth, at the time of the event, is traveling toward 
the Galactic Bulge, and the effect of the source velocity on the 
relative proper motion is suppressed by the distance factor).
This high proper motion yields linear velocities of 
$25-150\, {\rm km}\, {\rm s}^{-1}$.
Also the allowed directions of the lens proper motion lie between 
$90^\circ$ and $180^\circ$ relative to the Galactic North
(\ie, between Galactic East direction and Galactic South direction), 
this, in connection
with high proper motion yields high probability of
a substantial component perpendicular to the disk. 
This picture is more 
consistent with Galactic thick disk kinematics rather than thin disk.
Thus, in order to impose priors on the lens 
parameters, we assume that the lens belongs to the thick disk 
population. 
This is a less constraining assumption than thin disk, as thick disk 
allows for grater velocity dispersions.

We follow the approach of \citet[\S5.1]{batista11}, but using 
parameters suitable for a thick disk.
In the exponential disk we use scale height $=0.6\, {\rm kpc}$ and scale 
length $=2.75\, {\rm kpc}$. The probability density of the lens mass 
is assumed to be proportional to $M^{-1}$. We expect the velocity 
dispersion of the thick disk stars to be 
$(40,55)\, {\rm km}\,{\rm s}^{-1}$ in the North and East Galactic directions
respectively. The expected mean velocity is $(0,200)\, {\rm km}\,{\rm s}^{-1}$,
where we account for asymmetric drift of the stars of $20\, {\rm km}\,{\rm s}^{-1}$ 
behind the Galactic rotation.
The expected velocity of the Galactic Bulge sources is zero and its
velocity dispersion is $(100,100)\, {\rm km}\,{\rm s}^{-1}$.

From the disk model we assign prior probability density for certain values 
of lens distance, mass, and lens-source relative proper motion 
($D_l$, $M$ and $\bm{\mu}$). We translate these priors to microlensing
parameters ($t_\e$, $\theta_\e$, $\bm{\pi}_\e$) using the Jacobian 
derived by \citet{batista11}:
\begin{equation}
j_{\rm gal} =\left\lVert\frac{\partial(M,D_l,\bmu)}
{\partial(\theta_\e,t_\e,\bpi_\e)}\right\rVert =
\frac{2\pi_{\rm rel} M\mu^2}{t_\e \theta_\e\pi_\e^2}\frac{D_l^2}{{\rm AU}}
\end{equation}
where $M=\theta_\e/\kappa \pi_\e$, 
$D_l = \mathrm{AU}/(\pi_\mathrm{rel}+\pi_s)$,
$\pi_\mathrm{rel}=\theta_\e \pi_\e$, and
$\mu=\theta_\e/t_\e$. 

\subsection{MCMC results}
The results from MCMC are presented in Figure~\ref{f:chain-nat}, where
likelihoods obtained from the chain are projected onto the number
of 2-d planes, one for each pair of parameters. 
Those likelihoods are weighted by the priors
as calculated above. 
We present plots in a ``typically expected'' scale for each parameter 
to show how well each of them was measured. 
We also provide insets to show
the detailed shape of the likelihood contours (in under-diagonal panels). 
The diagonal panel for a given parameter 
shows the posterior likelihood marginalised along all other dimensions.
More exact numerical values of parameters from the region of maximal
likelihood can be found in Table~\ref{t:params}.

To preserve clarity, the described Figure 
does not show symmetric solutions 
that originate from the static binary ecliptic degeneracy, 
which is discussed in Section~\ref{s:degeneracies} and 
given by Equation (\ref{e:ecliptic}). 
These can be seen in the right panel of 
Figure~\ref{f:pien-thetadot-chain} projected onto the
$\pi_{\e,N}$-$\gamma_\perp$ plane.






\section{Physical parameters}
\label{s:params}

\subsection{Source star}

\begin{figure}
\plotone{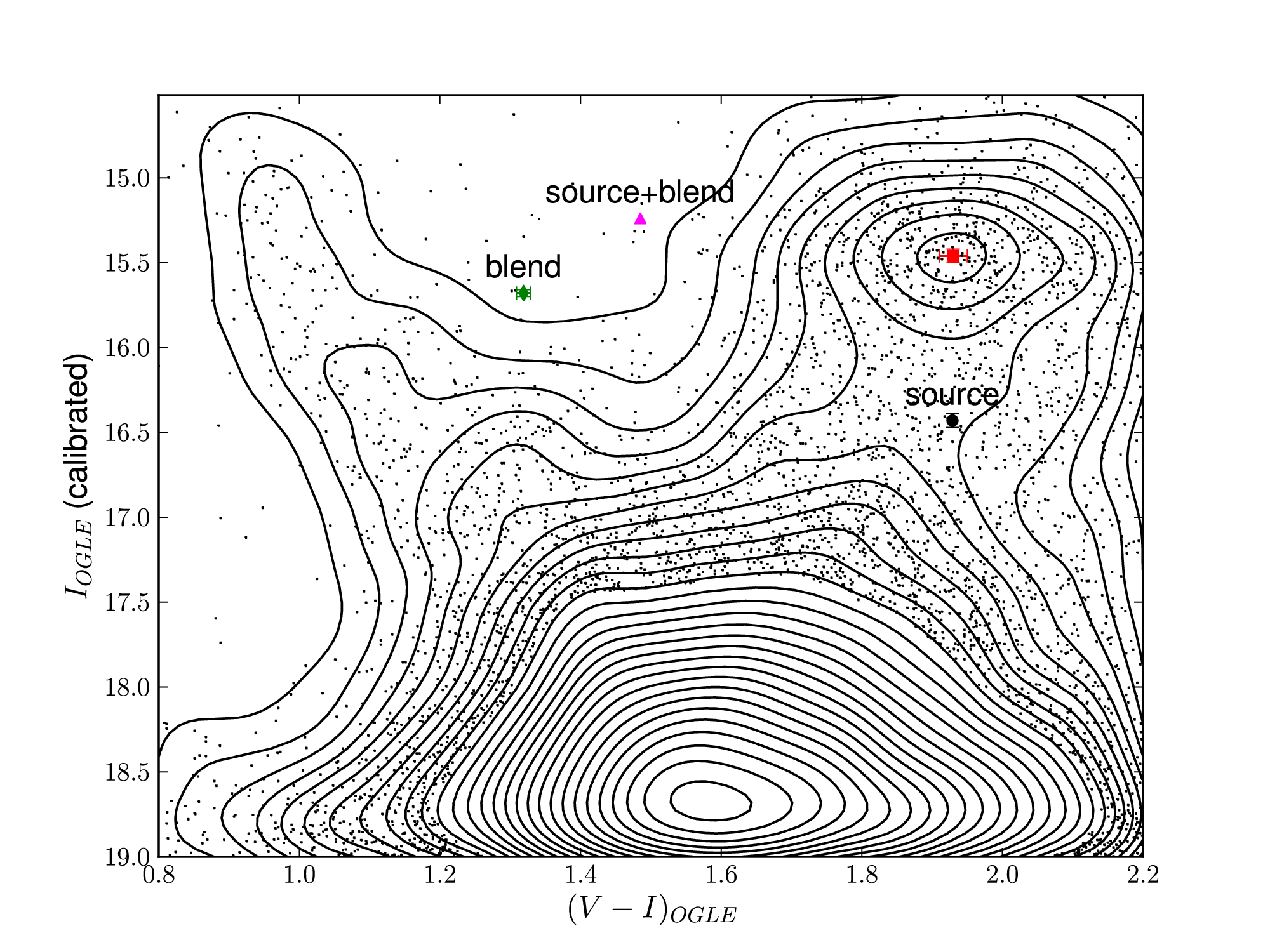}
\caption{Color-magnitude diagram for OGLE-III data. The square marks the 
position of the red-clump centroid. The circle is the brightness and 
color of the microlensed source star. The diamond is the light lying within 
the aperture that was not magnified (a blend), and the triangle is the sum 
of the blend and the source, as seen well before and after the 
microlensing event.\label{f:cmd}}
\end{figure}

\begin{figure}
\epsscale{.86}
\plotone{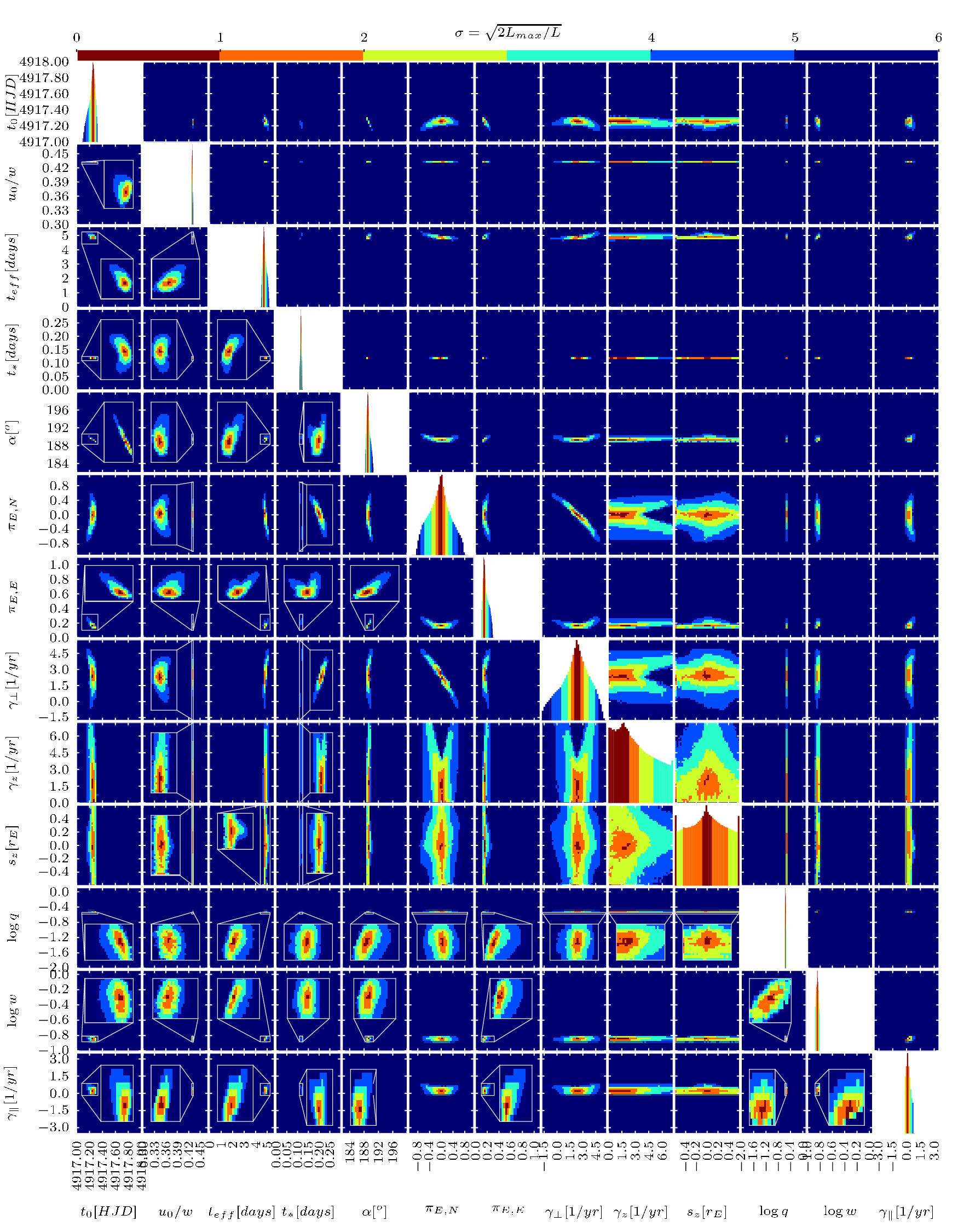}
\epsscale{1.0}
\caption{2-d projection of likelihoods in 13-dimensional space of the
microlensing parameters, of which the first 10 are the MCMC parameters
and the last 3 are the grid (\ie\ kept fixed during the MCMC
procedure) parameters. We present plots in typical (or natural) scale for each 
parameter for easy assessment of how well each of them is measured.
Over-diagonal panels are the same as under-diagonal ones except 
that the latter contain insets.
Diagonal panels show the likelihoods of one parameter (displayed on 
the corresponding horizontal axis) marginalized over all other 
dimensions. The height of the plots corresponds to the 7-$\sigma$
difference. Likelihoods shown are weighted by priors as described in \S\ref{s:priors}.
\label{f:chain-nat}}
\end{figure}

From OGLE-III photometric maps for the Galactic Bulge 
(\citealt*{udalski08}; Szyma{\'n}ski \etal\ 2011 in prep.) we construct a color magnitude 
diagram (CMD) of stars within $4'$ around the location of the event 
(Figure \ref{f:cmd}). The position of the centroid of the Red-Clump 
Giant stars on the CMD is derived to be:
\begin{equation}
\left( (V-I), V , I \right)_{RC,OGLE} = 
\left(  1.93 \pm 0.02, \; 17.39 \pm 0.04, \; 15.46 \pm 0.04 \right).
\label{e:redclump}
\end{equation}

We take the intrinsic centroid of the red clump (RC) to be 
$M_{I,RC,0} = -0.25 \pm 0.05$ \citep{bennett10}, 
and the color $(V-I)_{RC,0} = 1.08 \pm 0.06$ after 
\citet[\S4.5]{bensby10}, we calculate $M_{V,RC,0} = 0.83 \pm 0.08$
 and assume the distance to the Galactic Center 
to be $8.0 \pm 0.3$ kpc \citep{yelda10}. 
The location of the observed microlensing event is 
$(l,b)=(1^\circ 33' 29'', -3^\circ 49' 21'')$ in Galactic coordinates. 
When we take into account the inclination of the Galactic Bar we infer 
that the RC stars in the field are slightly closer than Galactic 
Center. From \cite{nishiyama05} we find a distance modulus smaller 
by $0.05$ mag. This leads to a RC distance modulus 
of $14.47 \pm 0.08$, which yields estimates of 
the reddening along the line of sight.
\begin{equation}
\left( E(V-I), A_V , A_I \right) = 
\left(  0.85 \pm 0.06, \; 2.09 \pm 0.12, \; 1.24 \pm 0.10 \right).
\label{e:reddening}
\end{equation}

\noindent This gives $R_{VI} = A_V/E(V-I) = 2.4$ which is comparable to 
other estimates of the extinction law toward the Bulge 
\citep[eg.][]{bennett10}.

We assume that the microlensing source is located behind the dust 
causing this reddening. Applying this average extinction to 
the source brightness obtained directly from the microlensing fit 
to the calibrated OGLE data:
\begin{equation}
\left( (V-I), V, I \right)_{OGLE} 
= \left(  1.929 \pm 0.002, \; 18.36 \pm 0.04, \; 16.43 \pm 0.04 \right)
\label{e:source_mag}
\end{equation}
\noindent we obtain the dereddened color and brightness of the source star:
\begin{equation}
\left( (V-I), V, I \right)_{0,OGLE} 
= \left(  1.08 \pm 0.06, \; 16.27 \pm 0.13, \; 15.19 \pm 0.11 \right)
\label{e:source}
\end{equation}

\subsubsection{Source radius}
\label{sec:theta_star}

Using color-color relations from \cite{bessell88} 
we infer that the source is a giant star of the spectral type K 
and we find $(V-K)_0 = 2.50 \pm 0.15$.
We assume solar metallicity for the source star. Then from 
\cite{houdashelt00}, we find that the temperature for a cool 
solar-metallicity giant star with $V-I=1.08$ should be about 
$4650\, {\rm K} \pm 100\, {\rm K}$. This is in agreement with \cite{ramirez05} 
who give the temperature equal to $4600\, {\rm K} \pm 100\, {\rm K}$.

To find the value of the surface gravity we use the \cite{berdyugina94} 
empirical relation for G-K giants and subgiants given by:
\begin{equation}
\log g = 8.00 \log T + 0.31 M_{0,V} + 0.27 [Fe/H] 
- 27.15,\; \textrm{with an accuracy of} \pm 0.25\, \mathrm{dex}
\label{e:logg}
\end{equation}
\noindent where we use
$M_{0,V} = 1.80 \pm 0.10$ to obtain $\log g = 2.71 \pm 0.26$.

\cite{kervella04} calibrated the color-brightness relations for 
giants. We take their relation (14) with dereddened visual 
magnitude of the source star $V_{0} = 16.27 \pm 0.13$
and $(V-K)_0$ color found above.
The resulting angular radius 
of the source is 
\begin{equation}
\theta_{*} = ( 4.45  \pm 0.40 )\, \mu\textrm{as.}
\label{e:thetastar}
\end{equation}

\subsubsection{Limb darkening coefficients}
\label{s:limb}

Knowing the temperature of the source star and its surface gravity 
we find the values of the linear limb-darkening coefficients. We use 
\cite{claret00} Table 30 with the assumption of solar metallicity 
and turbulence velocity of $2 \, \rm{km}\,\rm{s}^{-1}$. Uncertainties on these 
parameters have a very minor influence on the values of limb-darkening 
coefficients, with uncertainties in temperature being dominant factor.
For bands in which the observations have been performed we find:
\begin{equation}
u_V = 0.795, \quad u_R = 0.716, \quad u_I = 0.618, \quad u_H = 0.429 .
\label{e:limbu}
\end{equation}

For calculations of the magnification of the limb-darkened source 
we use coefficients in the form of \citet[\S3.1]{afonso00}. 
Their linear limb-darkening coefficient $\Gamma$ is related to 
commonly used coefficient $u$ with this relation: 
$u = 3\Gamma/(\Gamma+2)$. Thus, we have:
\begin{equation}
\Gamma_V = 0.721, \quad \Gamma_R = 0.627, \quad \Gamma_I = 0.519 ,
                  \quad \Gamma_H = 0.334 .
\label{e:limbgamma}
\end{equation}

\subsection{The source trajectory curvature degeneracy}
\label{s:piEN_omega}

There is a substantial degeneracy in the microlensing model between
the angular velocity of the lens in the plane of the sky 
($\gamma_\perp$) and the North component of the microlensing parallax
($\pi_{\e,N}$), which can be seen clearly on Fig. \ref{f:chain-nat} 
(the intersection of row 8 and column 6).

This degeneracy can be predicted to be present in a wide 
range of microlensing events. It is a purely geometrical effect --
both rotation of the lens axis and parallax motion of the Earth 
have similar effects on the apparent source trajectory in the lens plane.
This modification in both cases could be described, to the first order,
as a curvature of the source trajectory. Hence the curvature needed 
to explain observed changes in magnification can be a linear
combination of both effects.

\begin{figure}
\plottwo{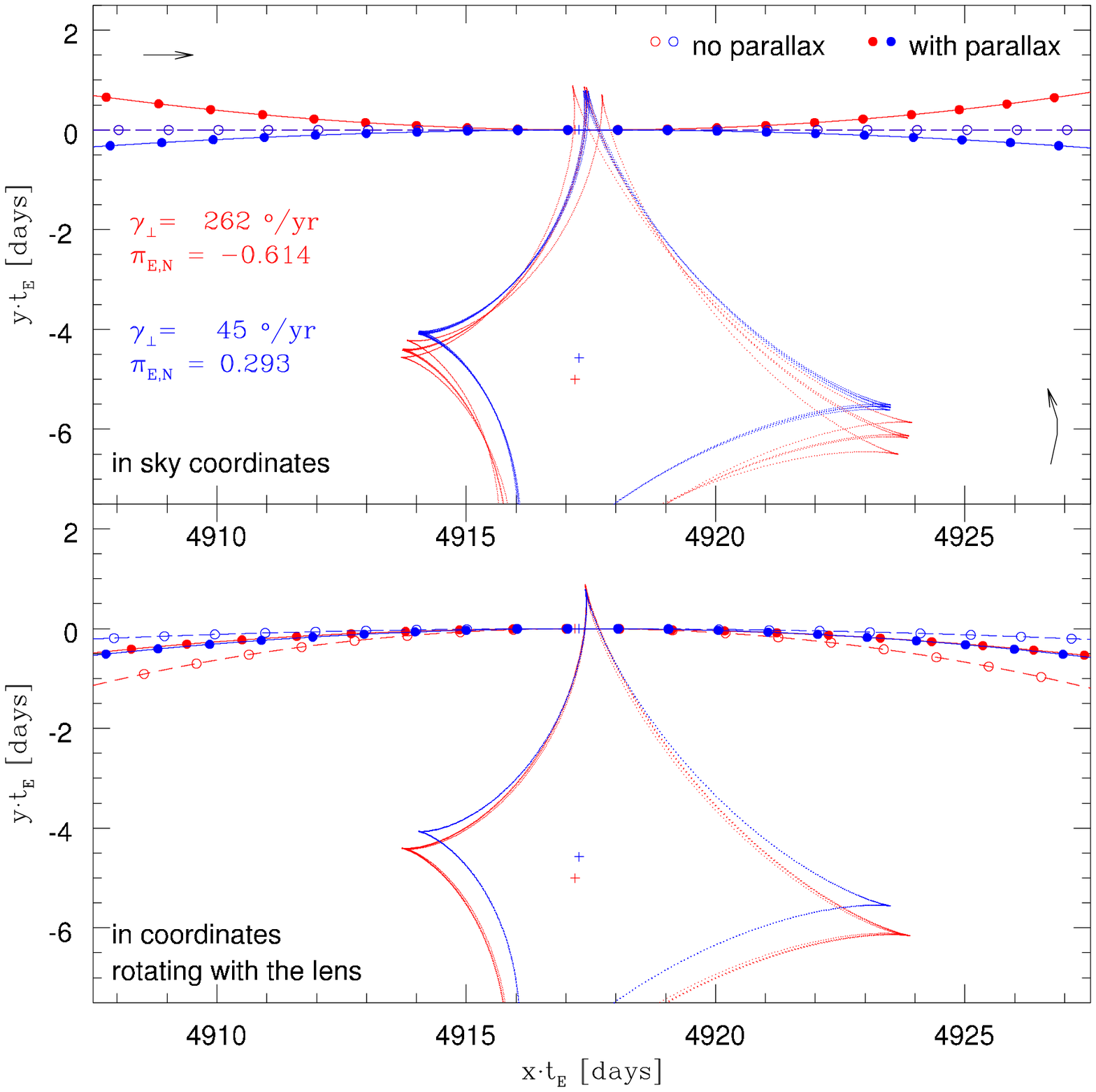}{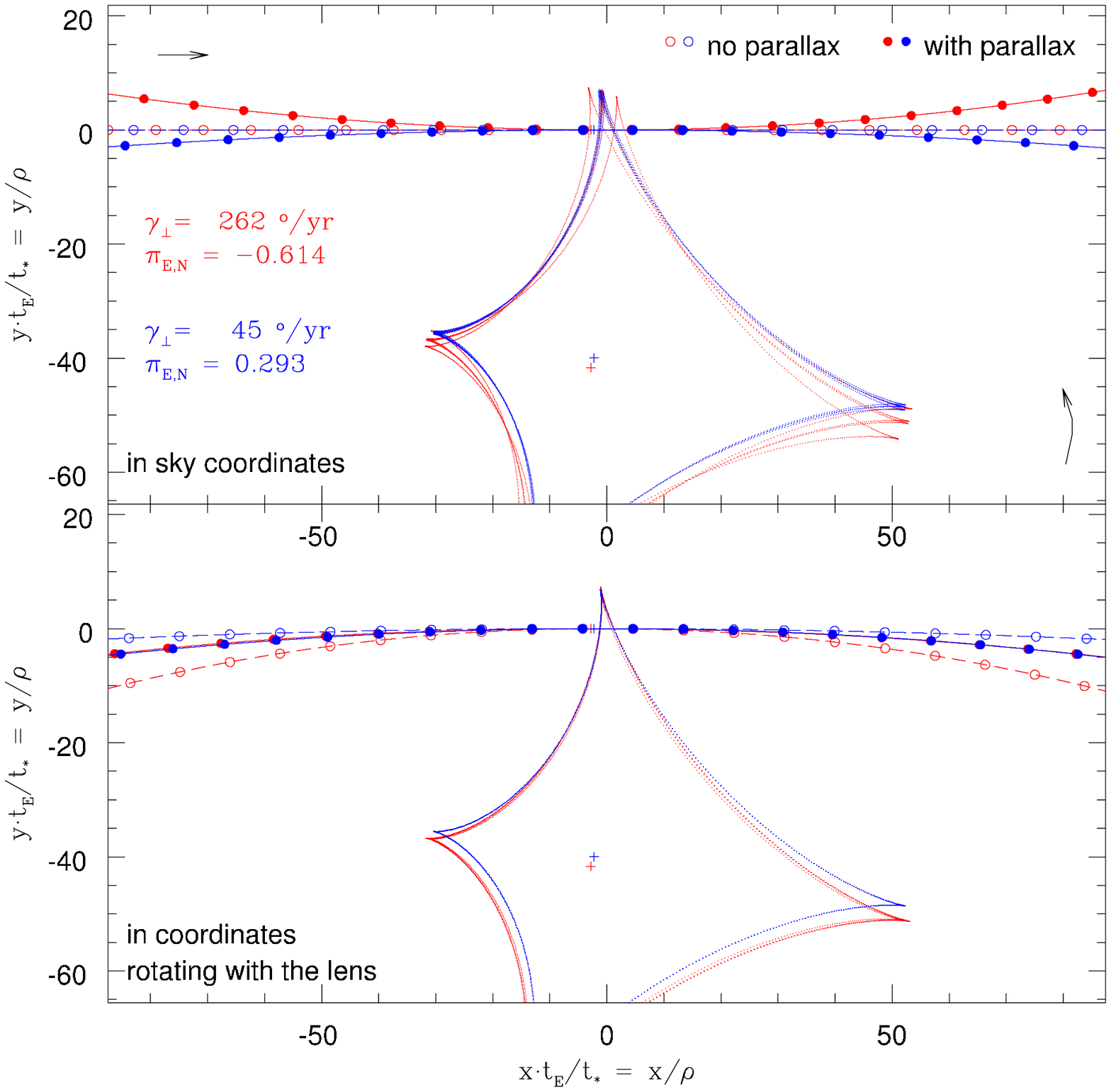}
\caption{Illustration of the `curvature' degeneracy (\S\ref{s:piEN_omega}): 
two microlensing 
model trajectories (red and blue) with extremely different values 
of $\pi_{{\rm E},N}$ and angular velocity ($\gamma_\perp$) lead to similar 
source-lens relative trajectories (filled circles in lower right panel) 
and thus similar goodness of fit.
Left panels show model trajectories in coordinates normalized 
by the timescale of a particular model. Right panels show the same
trajectories but with coordinates normalized by the size of the source
in each model.
Top row shows the projection of the trajectories and rotating 
caustics on the plane of the sky, while bottom row shows
projection of the same
trajectories on the plane rotating with the lens.
(Caustics are plotted at the times of caustic crossings and 4 days 
before and after).
Filled dots show the source positions once a day and open circles
show the source positions at the same points in time but
with the parallax shift subtracted as if there were no acceleration
of the motion of the Earth.
In summary, the different combinations of the effects of parallax and lens 
rotation can lead to the same projected source trajectory.
\label{f:pien-thetadot-models}}
\end{figure}

\begin{figure}
\plottwo{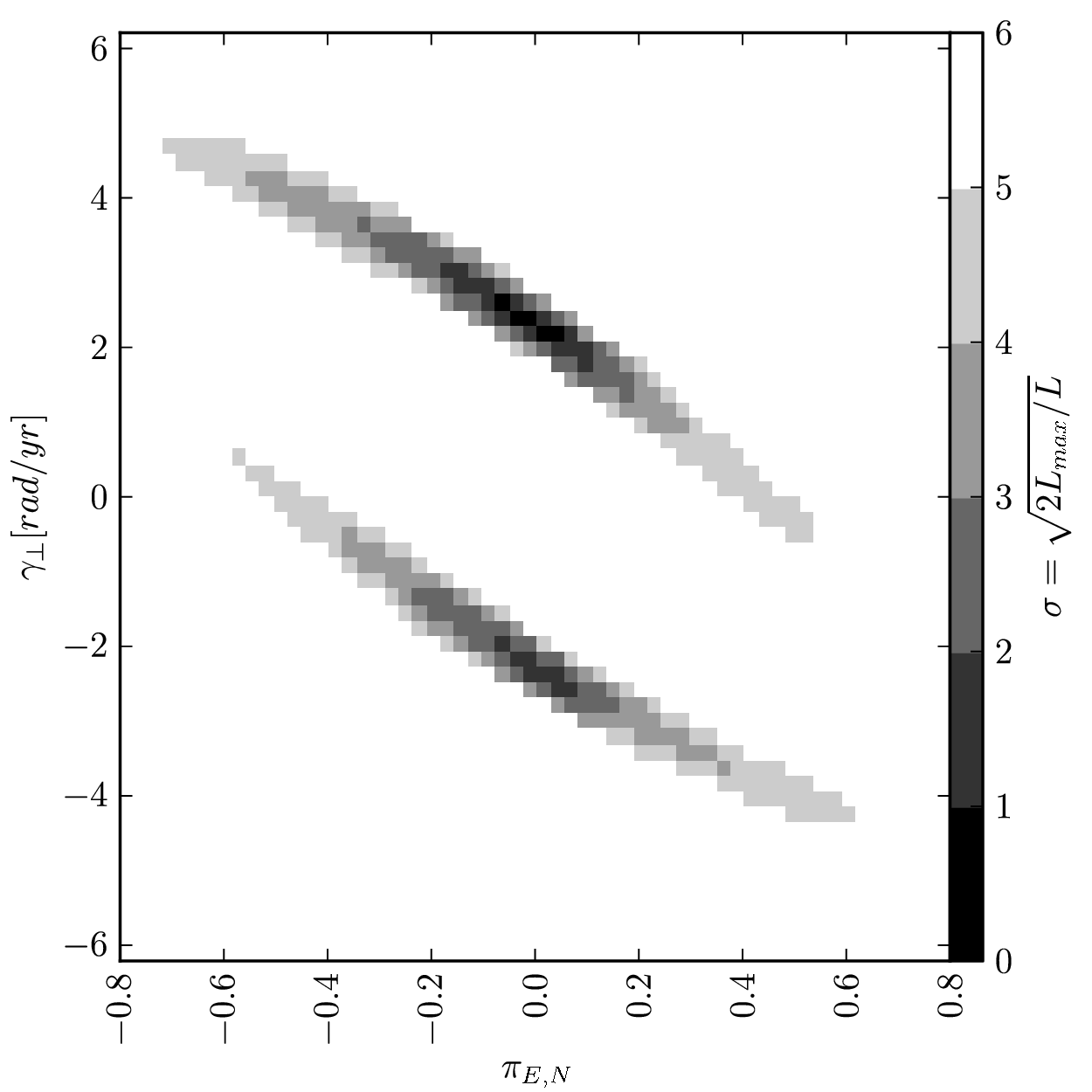}{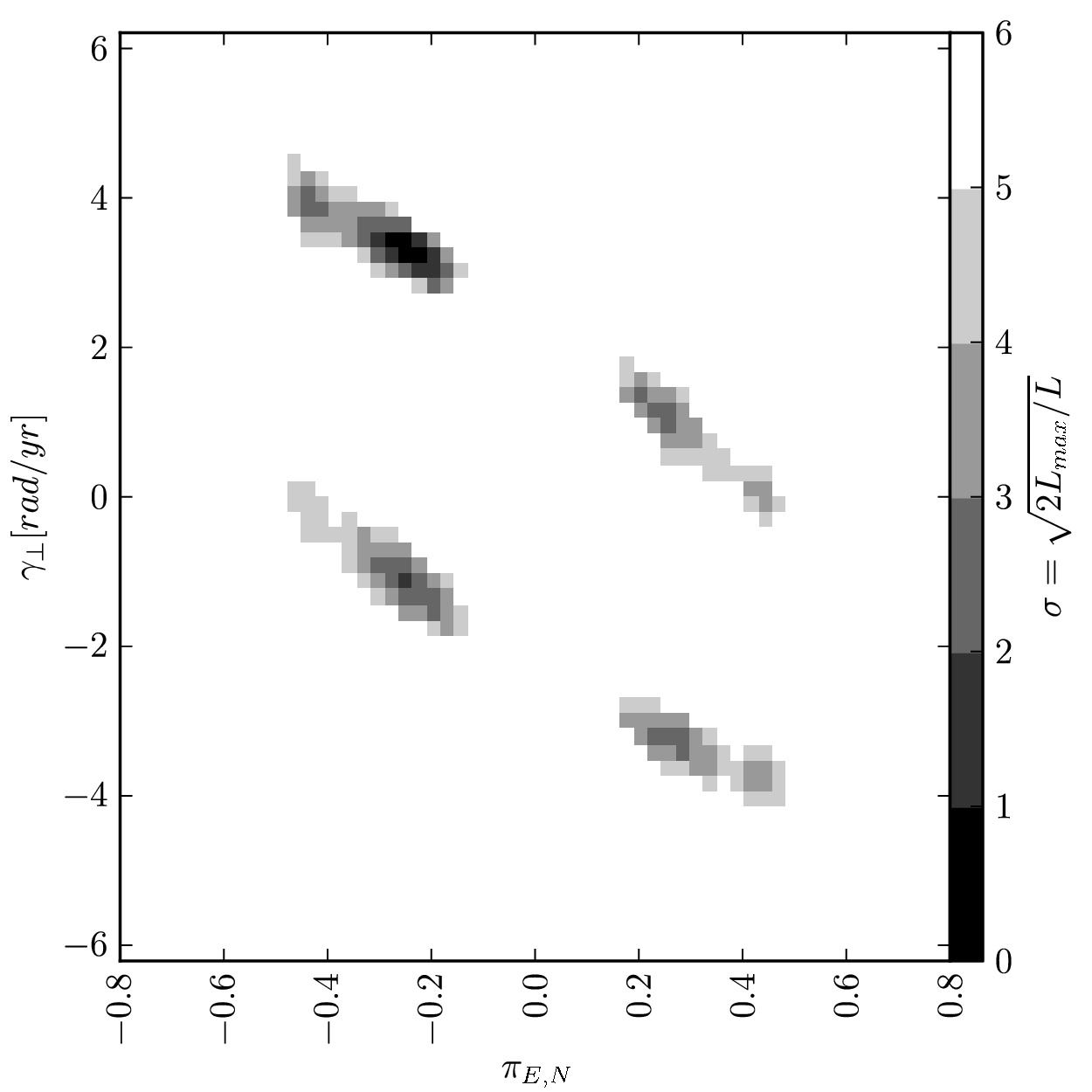}
\caption{Projection of likelihoods onto $\pi_{{\rm E},N}$ and
$\gamma_\perp$ plane. The left panel shows the full MCMC chain for 
solutions with positive $u_0$ (top region) and negative $u_0$
(bottom region). Links that survived a consistency check with 
the theoretical isochrone (\S\ref{s:cmd-cut}) are shown in
the right panel.
\label{f:pien-thetadot-chain}}
\end{figure}

Here we stress the importance of this degeneracy.
Our model minimization shows that it could be very severe and 
have a significant impact on the final value of the distance 
to the lens and its mass.
Figure \ref{f:pien-thetadot-models} shows two models,
both fitting the observed light curve well, 
with extremely different values of $\pi_{\e,N}$ and $\gamma_\perp$, 
which lead to a very similar effective trajectory of the source on
the plane of the lens and finally to similar goodness of fit.
Masses and distances derived from these solutions differ by a factor of 2.
Note, that both trajectories overlap more closely when plotted
in units of source radius rather than time. 

In the general case of microlensing events observed toward the Galactic 
Bulge and with the parallax signal measured, it is very common that the
$\pi_{\e,E}$ component is well measured and $\pi_{\e,N}$ has substantial 
uncertainty. Looking at Figure 3 in \citet{park04} one sees 
that, if we decompose the parallax vector ($\bm{\pi}_\e$) into 
components parallel ($\pi_\parallel$) and perpendicular ($\pi_\perp$) 
to the direction of the Earth acceleration (near the peak 
 of the event), actually it is the $\pi_\parallel$ that is well 
measured and the big uncertainties are in $\pi_\perp$.
Since the Earth acceleration vector 
most of the year lies near the East-West direction, the
uncertainty in $\pi_\perp$ usually translates to uncertainties
in $\pi_{\e,N}$.

For this event, the caustic crossing occurred on 
27 March 2009, so very close to the vernal equinox when the apex of the 
Earth motion was almost directly toward the Galactic Bulge. That is why 
the direction of Earth acceleration was very close to East, so
$\pi_{\e,N}$ $(\approx \pi_{\e,\perp}$ in this case) is not well 
constrained.

Because of this ``curvature'' degeneracy we are unable to 
accurately determine the distance and the mass of the lens from 
the microlensing fit only.
We need to use other known system parameters such as 
observed magnitudes and colors of the lens (where we assume
that all blended light is comming from the lens). 
We therefore restrict the 
set of solutions to those that are consistent with
theoretical color-color and color-magnitude relations.

\subsubsection{Choosing solutions consistent with the theoretical isochrones}
\label{s:cmd-cut}

We take a set of Y$^2$ (Yonsei-Yale) 
Isochrones\footnote{\url{http://www.astro.yale.edu/demarque/yyiso.html}}
\citep{demarque04} and from the microlensing fit to the $V$-, $I$- and 
$H$-band data we infer the magnitudes of the blend.
Each link in the chain of solutions gives slightly different 
values of the observed magnitudes of the blend 
but, for reader information, we quote the most common ones:
\begin{equation}
\left((V-I), I, (I-H)\right)_b = \left(1.316 \pm 0.01,\; 15.680 \pm 0.01,\;  1.409 \pm 0.05\right).
\label{e:blend_mag}
\end{equation}
We assume that
all non magnified light comes from the binary lens and that the 
components of the lens are main sequence stars. Given the mass ratio 
$q = 0.27$ from the microlensing fit we neglect the light input from 
the secondary and check the consistency of the mass 
of the primary, its distance, and its luminosity against the 
theoretical isochrones.

For initial tests we take the isochrone with solar metallicity and 
age. For each solution (link) we calculate the distance and 
mass of the primary and from the interpolated isochrone we find the 
theoretical magnitudes in $V$, $I$ and $H$ that it should have if 
seen with no reddening. 
Then, assuming the slope of the reddening curve $A_I/A_V=0.6$ and
$A_H/A_V=0.17$, we find the amount of interstellar reddening 
in $V$-band ($A_V$) that we should subtract in order to shift the
observed magnitudes closest to the theoretical values. We throw out all
solutions requiring negative reddening or more reddening than the total 
amount seen toward the Galactic Bulge source. We also add gaussian 
weights to each individual link depending on how close it could be placed
to the theoretical magnitudes with respect to the error bars, and a weight 
corresponding to the value of the reddening assuming that it should be
about $1 \pm 0.5 \, \rm{mag}\,\rm{kpc}^{-1}$ in $V$.

The set of solutions with highest weights lies near 
$\dd A_V / \dd D = 1.4 \, \rm{mag}\,\rm{kpc}^{-1}$,
distance $\approx 1$ kpc and mass $\approx 1 \, M_\Sun$
for the primary.
From the most probable values of parallax parameters ($\pi_{\e,N} \approx -0.2$
and $\pi_{\e,E} \approx 0.2$) we see that the likely direction of
the transverse velocity 
of the lens relative to the line of sight is perpendicular 
to the Galactic plane. The value of the relative proper motion taken
from the equation $\mu = \theta_\e / t_\e$ is near  $12 \, \rm{mas}\,\rm{yr}^{-1}$,
which at the distance of $1$ kpc translates to a linear transverse 
velocity of $\sim 60 \, \rm{km}\, \rm{s}^{-1}$.
These, further constrained kinematic parameters, strengthen our prediction
(from \S\ref{s:priors_lens}) that it is more likely that the lens
belongs to a thick disk population.
Thus, for final isochrone consistency checks, instead of a Solar-like isochrone,
we choose one with an age of $10$ Gyr, a sub-solar metallicity 
($[Fe/H]=-0.5$) and an $\alpha$-enhanced mixture ($[\alpha/Fe]=0.3$) which
would be more typical for stars with this kind of kinematics.

We prepare a subset of our chain links in which every link has weight
added depending on how well it matches with the theoretical 
isochrone. 
The region of parameter space that coincides with
the magnitudes from the isochrone, has:
\begin{equation}
D_l = 1.1 \pm 0.1 \, {\rm kpc} \quad
M_1 = 0.84 \pm 0.03 \, M_\Sun \quad
\frac{\dd A_V}{\dd D} = 1.4 \pm 0.2 \, {\rm mag}\,{\rm kpc}^{-1}
\label{e:mass_dist}
\end{equation}

\subsubsection{Choosing solutions consistent with the spectroscopic mass}
\label{s:spec-cut}

Even if some assumptions in the procedure described in the previous 
section happen to be incorrect, we will see this clearly from the first
spectrum taken of the object. The mass of the primary could be 
estimated from the spectrum, and it will be easy to redo the 
filtering of solutions with this additional information. 

In either case, the selected set of solutions could be subsequently 
tested against the observed RV curve to 
see if the orbital parameters derived from the microlensing fit are 
consistent with it.

\section{Discussion}
\label{s:discussion}

\subsection{Degeneracies}
\label{s:degeneracies}

Nine of the microlensing parameters are known from the fit with 
very high precision. However, there are two pairs of parameters for
which the values are imperfectly measured (see Figure \ref{f:chain-nat}).
The first, obvious, pair is the position and velocity along the 
line of sight ($s_z$ and $\gamma_z$), which do not have direct 
effects on the observed magnification. They are only limited by the
requirement of $e<1$ and by the shape of the Keplerian orbit 
projected on the plane of the sky -- to be precise, only by the 
segment of the orbit that was covered
by the lens components during the high magnification period.
The other pair of parameters that was not measured well are 
$\pi_{\e,N}$ and $\gamma_\perp$, which are degenerate as was described
in Section \ref{s:piEN_omega}.
We partially break this degeneracy into discrete regions by choosing 
only those solutions that are in agreement with other information
we have on the event (sections \ref{s:cmd-cut} and \ref{s:spec-cut}).

Because the mass of the lens depends on the magnitude of the vector 
$\bm{\pi}_\e = (\pi_{\e,N}, \pi_{\e,E})$ it does not change when the sign 
of the $\pi_{\e,N}$ component changes. This leads to 2 regions along the 
$\pi_{\e,N}$-$\gamma_\perp$ degeneracy that yield the same mass.

There is one more degeneracy from which our model suffers, \ie, the
orbiting binary ecliptic degeneracy (see \S \ref{a:orbit2d}).
We note that the ecliptic degeneracy happens when the direction of the
Earth acceleration is constant: the one obvious case, is when the 
source lies on the ecliptic
(hence the name of this degeneracy), but the other is when the timescale of the
event is much shorter than the characteristic timescale of the changes of
the direction of acceleration. This is more likely to occur if the
microlensing event happens when Earth is moving toward or away from the
source. 

This is the case in OGLE-2009-BLG-020, for which the peak 
magnification happened near the vernal equinox. The Earth acceleration at that
time is directed toward East and its direction varies very slowly 
in time. 
Note also that at that time $\pi_{\e,\perp} \approx \pi_{\e,N}$ so,
following (\ref{e:orb_ecl_deg}), from any set of parameters we can
obtain an analogous set by changing:
\begin{equation}
(u_0,\alpha_0,\pi_{\e,N},\gamma_\perp)
\rightarrow -(u_0,\alpha_0,\pi_{\e,N},\gamma_\perp)
\label{e:ecliptic}
\end{equation}

In Figure \ref{f:pien-thetadot-chain}, we see 
the projection of our set of solutions 
on the plane of $\pi_{\e,N}$ and $\gamma_\perp$. The two allowed regions
correspond to positive $u_0$ (on top) and negative $u_0$ (bottom)
solutions. The left panel shows the full chain while
the right panel shows the subset of links that are consistent with the
isochrones chosen in Section \ref{s:cmd-cut}.
Each of the continuous regions from the right panel is split into 2 regions
(for positive $\pi_{\e,N}$ and negative $\pi_{\e,N}$), giving 4 separate 
regions in total. 

Note that none of the
degeneracies described in this section is exact, this leads to 
different likelihoods between the 4 regions.

\subsection{Symmetries}
Changing the sign of $\gamma_\perp$ without changing the other 
components of the velocity flips the orbit, which manifests itself as a
change of some Euler angles: $\Omega_{node} \rightarrow -\Omega_{node}$
and $i \rightarrow \pi - i$.

As microlensing cannot directly measure position and velocity along 
the line of sight every solution has its copy mirrored by the 
plane of the sky, \ie, $s_z \rightarrow -s_z$ and 
$\gamma_z \rightarrow -\gamma_z$. For orbit elements it means:
$\Omega_{node} \rightarrow \pi-\Omega_{node}$ and
$\omega_{peri} \rightarrow \omega_{peri}-\pi$.

Without loss in generality we perform all calculations 
assuming $\gamma_z>0$ at the chosen time ($t_{0,\kep}$).
Radial velocity measurements will have information on the sign of the
radial velocity -- we can then simply mirror our set of solutions
using the above prescription.

\subsection{Test of the microlensing solution with Radial Velocity}
\label{s:test}

Every link in the MCMC chain represents a complete set of parameters
of the binary lens. 
It yields not only the mass, distance and separation
in physical units, but also all Keplerian parameters of the orbit.
Thus we can calculate the RV curve for any period of the binary.

Assuming the observations of the RV curve of the primary are taken,
we can assign to every link the likelihood that it is consistent with 
those data. We derive the radial velocity for any time the data were
taken and, allowing the systemic velocity of the center of mass of the binary
to vary, we calculate the likelihoods for every point.
In this way we construct a new set of links weighted by both the microlensing 
light curve and the radial velocity curve. 
This new set of solutions yields new, most probable, values of 
all lens parameters, which may or may not coincide with the values
derived from the microlensing only solution. This would be a test of
the microlensing solution. 

If the new values of the parameters lie outside the 3-$\sigma$ limits
of the microlensing solution we can say that our method failed.
By contrast, if solutions, which are consistent with the RV curve,
lie near the best-fit values we obtained from microlensing, we can 
not only believe our solution, but we can also read off all parameters 
of the binary, which are not given by one or the other method alone.
For example, the radial velocity curve will give us the period and systemic
radial velocity, which we cannot read from the microlensing 
light curve alone. However, microlensing will yield
the inclination, orientation on the sky and the 2-d velocity of the binary
projected on the plane of the sky.

\subsubsection{Example of the test}
\begin{figure}
\epsscale{.86}
\plotone{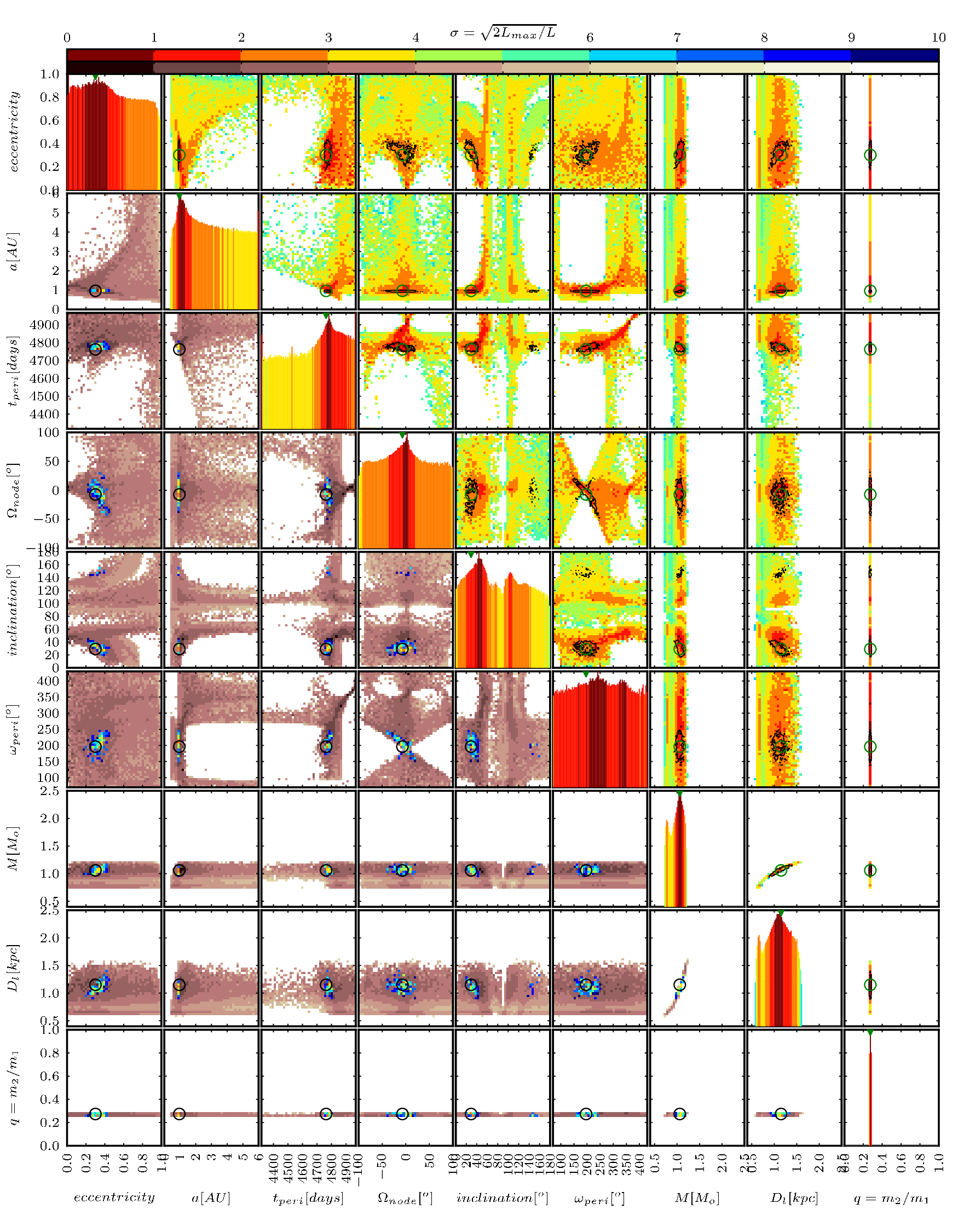}
\epsscale{1.0}
\caption{Likelihoods in the space of Kepler parameters shown in colors
in the over-diagonal panels, and as a gray-scaled background 
in the under-diagonal panels. 
The figure illustrates also how observed RV curve would recover 
most probable values of parameters when compared with our set
of solutions (\S\ref{s:test}). One exemplary solution is chosen from the MCMC 
simulation as the ``true'' underlying binary parameters 
and is marked with open circles. 
Likelihoods of values of Kepler parameters in agreement with both
the microlensing light curve and the radial velocity curve are shown in color
in the under-diagonal panels (and as a 10-$\sigma$ black contour on
the over-diagonal panels). 
\label{f:kepler}}
\end{figure}

We illustrate this test in Figure \ref{f:kepler}. The Figure shows
(in the background) the likelihoods 
derived from microlensing solution projected onto 2-d planes of 6 orbital
parameters, the mass, distance and mass ratio  
(this is a subset of the whole chain chosen using the method described in 
\S\ref{s:cmd-cut}).
We take one, exemplary set of binary parameters and simulate the 
radial velocity curve consisting of 15 measurements taken from 
March to October 2011 and 15 measurements in a similar period in 2012.
We then test our chain against this RV curve assuming  
a measurements precision of $0.5\, {\rm km}\, {\rm s}^{-1}$. 
The solutions with the highest joint likelihoods (from microlensing and 
RV curve fitting) are overplotted on each panel (in color on under-diagonal
panels, and as 10-$\sigma$ contour on over-diagonal panels). 
The initial set of parameters we chose to generate the RV curve is 
marked with open circles. 

We see that all orbital parameters of the ``seed'' set of parameters
are retrieved by the RV comparison process (as the links selected
to be consistent with RV curve lie near the circles). This shows
the selective power of the radial velocity data, which can prove 
whether the 
microlensing solution is wrong or strongly suggest that it is right.

\subsubsection{Useful information for RV observations}
The selective power of the RV curve holds when the observations 
are taken through at least one period of the binary. 
We have chosen a 2 year span of observations since the period 
we derive from our solutions is between 200 and 700 days (2-$\sigma$ limit).

The binary equatorial coordinates are 
($18^h 04^m 20^s.99$, $-29^\circ 31' 08''.6$, $J2000.0$). 
The finding chart can be found on the OGLE EWS webpage -- the link
is given in the footnote in Introduction.
The binary is blended with the ``microlensing source'' which, as the
CMD in Figure \ref{f:cmd} suggests, is a Galactic Bulge giant.
The observed magnitudes of the source are given by Equation (\ref{e:source_mag}). 
The binary has a mass ratio of $0.272$, so assuming both components
are main-sequence stars, the majority of the light is coming from the primary.
Observed magnitudes of the binary are given by Equation (\ref{e:blend_mag}).
The binary is $1.3$ magnitude brighter in $V$ and $0.8$ mag brighter in $I$
than the giant, however they are of similar brightness in $H$.

Since the binary is located in the Galactic Disk, we anticipate it will be clearly 
separated in the velocity space from the blended Bulge giant.
The mass and the distance of the primary, which we predict from the
comparison of the light coming from the binary with the theoretical 
isochrone, is given by Equation (\ref{e:mass_dist}) in \S\ref{s:cmd-cut}.
The radial velocity amplitude of the primary is expected to be
$5 \pm 1 \, {\rm km}\,{\rm s}^{-1}$.

\section{Conclusions}
\label{s:conclusions}

The binary star that manifested itself in the microlensing event 
OGLE-2009-BLG-020 is the first case of a lens that is close enough 
and bright enough to allow ground-based 
spectroscopic follow-up observations.
This makes it a unique tool to test the microlensing solution.

We derive lens parameters using the same method by which the majority of planetary 
candidates discovered by microlensing are analyzed. 
We detect a signal from the orbital motion of the lens in the microlensing light
curve. This signal, as well as our measurements of the orbital parameters of
the binary lens, can be confirmed or contradicted by future 
observations. We propose a test in \S\ref{s:test}.

Combining the microlensing solution with the radial velocity curve
will yield a complete set of system parameters including 3-d
Galactic velocity of the binary and all Keplerian orbit elements.

This work undertakes an effort to establish a uniform microlensing 
notation, extending work of \citet{gould00} by 
including the full set of orbital elements of the binary lens
(Appendix \ref{a:notation}). 
We also summarize all known microlensing symmetries and degeneracies.

The method of deriving orbital elements from the 6 phase-space
coordinates, used to parametrize microlensing event,
is described in Appendix \ref{a:jacobian}.
The Fortran codes we use for transformation of the microlensing parameters
to orbital elements and for deriving all quantities described in 
the Appendix \ref{a:jacobian} will be attached to {\it astro-ph}
sources of this paper, and will be published on the author's 
web page\footnote{\url{http://www.astronomy.ohio-state.edu/\textasciitilde{}jskowron/OGLE-2009-BLG-020/}}.


\acknowledgments

\noindent{\bf Acknowledgments:} We acknowledge following support:
JS: NASA grant 1277721 issued by JPL/Caltech and
Space Exploration Research Fund of The Ohio State University;
AG: NSF grant AST-0757888; AG, RP and SG: NASA grant NNX06AF40G.

Work by SD was performed under contract with the California Institute
of Technology (Caltech) funded by NASA through the Sagan Fellowship
Program. 
The OGLE project has received funding from the European Research Council
under the European Community's Seventh Framework Programme
(FP7/2007-2013) / ERC grant agreement no. 246678 to AU.
The MOA project acknowledge support of the JSPS 20340052 and 
JSPS 18253002 grants.
Work by CH was supported by the grant 2009-0081561 of the National
Research Foundation of Korea.

This work was supported in part by an allocation of computing time 
from the Ohio Supercomputer Center under the project PAS0367.
We thank David Will for administering and maintaining the
computer cluster at the OSU Department of Astronomy, which was 
extensively used for the purpose of this work.

JS thanks Dr Martin D. Weinberg for helpful discussions.

This publication makes use of data products from the 
Two Micron All Sky Survey, which is a joint project of the University 
of Massachusetts and the Infrared Processing and Analysis 
Center/California Institute of Technology, 
funded by the National Aeronautics and Space Administration 
and the National Science Foundation.


\clearpage
\appendix

\section{Notation, Conventions, and Symmetries}
\label{a:notation}

\citet{gould00} argued that it would be useful to establish a ``standard''
system of microlensing notation to both ease comparison of microlensing
light-curve fits by different authors, and to facilitate easy entry
into the field by outsiders.  Since that time, a number of new effects
have been modelled, forcing the introduction of new parameters.  As was
previously the case for the smaller set of parameters listed by \citet{gould00},
this has resulted in the emergence of multiple systems of notation,
often specifically adapted to the problem at hand.

In this paper, we have for the first time, fit a  microlensing light curve
to parameters representing a complete Kepler orbital solution.
Thus, it is appropriate to revisit the question
of notation.  Indeed the urgency of developing a common set of not
only notation, but also conventions, is increased, because it is 
becoming increasingly difficult to compare microlens solutions of
different groups without a ``score card''.

To the extent possible, we would therefore like to
establish a ``standard system'' of notation and conventions.  
However, our fundamental goal is actually slightly less ambitious:
to establish a ``reference system'' of notation and conventions.
Then, even if various researchers do not adopt this system, they
can still specify how their system is related to this reference system,
which will then enable direct comparison of solutions carried out
in different systems, and also fairly direct translation of parameters
from one system to another.

Finally, as the list of parameters being fit grows, so does the
exposition of the meaning of these parameters, repeated in one
paper after another, with slight variation.  It will be more convenient
(and cheaper) if future authors can simply reference this appendix,
perhaps supplemented by a few words on how their system differs.

Our general approach will be to begin with the \citet{gould00} system, and
extend it to new parameters that have ``spontaneously'' appeared in
the literature.  The extensions will be guided first by establishing
a logically consistent system, and second (to the extent possible)
adopting the most popular notation previously developed.  A big
``logical'' consideration is to define parameters that
closely parallel observable quantities, so as to avoid introducing
unnecessary degeneracies into the fitting process, as would be the case for
some physically well-motivated parameters that are not directly constrained
by the data.

We will outline a notation system that includes 
3 ``basic'' and 3 ``higher-order'' point-lens event parameters, 
3 additional parameters required to
describe static binaries, 2 additional parameters required to describe
binary orbital motion in the plane of the sky, and 4 further
parameters to describe complete orbital motion.  In fact, one of these
last four (the angular Einstein radius $\theta_\e$) can logically
be (and is) introduced much earlier, although it is not essential as
fit parameter until the last stage.
That is, there are 15 parameters in addition
to flux parameters $(f_s,f_b)$, describing the source flux and blended
flux from each observatory.  We do not include here parameters for
source orbital motion (``xallarap''), nor 3-body  systems.

The definition of parameters necessarily requires that we specify
certain conventions.  We also attempt to make these logically
ordered and consistent.  Finally, there are two main classes of
degeneracies, one continuous and the other discrete.  We trace how
these two degeneracies ``evolve'' as additional parameters are added
to the description of the system.

\subsection{Point Lens Parameters: Basic}
\label{s:pointlensbasic}

For point lens events, the almost universally accepted 
three basic parameters are
\begin{equation}
(t_0,u_0,t_\e)\qquad \rm (point\ lens\ basic).
\end{equation}
These are the time of closest approach to the lens system ``center'',
the lens-source projected separation at that time (in units of the
Einstein radius), and the Einstein crossing time.  The meaning of
system ``center'' is obvious in the case of a point lens, but will
require generalization for more complicated systems.   Moreover,
even $t_\e$ will require more exact specification.  We retain these
three notations, but defer discussion of the generalizations of their
meaning until further below.  A derived parameter, which is sometimes
used as an independent fitting parameter in place of either $u_0$
or $t_\e$ is the ``effective timescale''
\begin{equation}
t_{\rm eff} \equiv u_0 t_\e.
\end{equation}

\subsection{Point Lens Parameters: Higher Order}
\label{s:pointlenshigh}

There are three higher-order parameters that can in principle
be measured for point-lens events, and these lead to a fourth
derived parameter.  The first two parameters are the ``vector microlens
parallax''
\begin{equation}
\bpi_\e \equiv (\pi_{\e,N},\pi_{\e,E})\equiv 
(\cos\phi_\pi,\sin\phi_\pi)\pi_\e.
\end{equation}
Here $\pi_\e={\rm AU}/\tilde r_\e$, where $\tilde r_\e$ is the Einstein
radius projected onto the observer plane and $\phi_\pi$ is the
direction of the lens motion relative to the source expressed
as a counter-clockwise angle north through east.  The mere introduction
of $\bpi_\e$ brings with it a large number of symmetries and questions
of convention, which will grow yet more complicated as binary lenses
come into play.  We therefore carefully delineate these in their
simpler form here.

First, it has become customary to adopt the ``geocentric framework''
\citep{an02,gould04a}, in which all parameters are measured
in the instantaneous frame that is at rest with respect to the Earth
at a specifically adopted time:
\begin{equation}
t_{0,\rm par}\qquad \rm (parameter\ reference\ time)
\end{equation}
which is {\it not} a fit parameter.  Note that the subscript stands
for ``parameter'' (not ``parallax'').  Thus, for example, $u_0$ is the
distance of closest approach of the source and the ``lens center''
(i.e., for point lenses, simply the lens) in this geocentric frame.
Similarly, $t_0$ is the time of this closest approach, and $t_\e$
is the time it would take to cross the Einstein radius if the 
lens-source relative motion were the same as it is as seen from the
Earth at $t_{0,\rm par}$.  Since the Earth velocity is constantly changing,
all of these parameters depend on the choice of $t_{0,\rm par}$.
Finally, the direction $\phi_\pi$ (but not the magnitude $\pi_\e$) of 
the parallax vector also depends on this choice.  In practice, for
point lenses, $t_{0,\rm par}$ is chosen quite close to $t_0$ so that
this is hardly an issue.  But the issue will become more important
for binary lenses.

Second, point-lens parallaxes are subject to 4 related ``degeneracies'',
which one might also call ``symmetries''.  To properly express these,
we introduce a different set of basis vectors 
$\bpi_\e = (\pi_{\e,\parallel},\pi_{\e,\perp})$, where
$\pi_{\e,\parallel}$ is the component parallel to the apparent
acceleration of the Sun (projected on the sky) in the Earth frame
at $t_{0,\rm par}$ and the pair ($\pi_{\e,\parallel},\pi_{\e,\perp}$)
is right-handed \citep[Fig.~3 of][]{gould04a}.  We also must define a convention
for the sign of $u_0$:
\begin{equation}
u_0>0 \quad \Leftrightarrow \rm (lens\ passes\ source\ on\ its\ right)
\end{equation}
as in \citet[Fig.~2]{gould04a}.  Then
\begin{description}
\item[1) $\pi_{\e,\perp}$ Degeneracy:] Typically, $\pi_{\e,\parallel}$
is much better determined  than $\pi_{\e,\perp}$ \citep{gould94} because
the former is determined at third order in time and the latter at fourth order
\citep{smith03}.  This can lead to elongated error ellipses in
the $\bpi_\e$ plane.  Only for sufficiently long events (or strong
parallax signal) is this degeneracy broken.
\item[2) $u_0$ Degeneracy:] In the limit that the Earth's acceleration
can be regarded as constant, there is a perfect symmetry between
$\pm u_0$ solutions, with only minor adjustment of other parameters
\citep{smith03}.
\item[3) Ecliptic Degeneracy:] In the limit of constant {\it direction
of acceleration} (as would be the case for a source
on the ecliptic) there is a perfect degeneracy
\begin{equation}
(u_0,\pi_{\e,\perp}) \rightarrow -(u_0,\pi_{\e,\perp})
\end{equation}
\citep{jiang04,poindexter05}.  Hence, sources lying near
the ecliptic (i.e., all Galactic Bulge sources) may suffer an approximate
degeneracy. We note that at times when Earth is moving  
toward or away from the Galactic Bulge (near equinoxes), the direction of Earth 
acceleration projected on the sky varies slowly in time. 
This significantly extends the effect of this degeneracy.
\item[4) Jerk-Parallax Degeneracy:] This is a discrete degeneracy
described in detail by \citet{gould04a}.
\end{description}

The third higher-order point-lens parameter is $\rho$, the radius of
the source in units of the Einstein radius.
Of course, $\rho$ is only rarely measurable in point-lens events
and is very frequently measured in binary events, and so is usually
called a ``binary-lens'' parameter, but it logically precedes the
introduction of binary events, so we include it here.  
The angular radius of the source $\theta_*$ can almost always be measured from
the instrumental color and magnitude of the source \citep{yoo04}.
When $\rho$ is also measurable, then one can measure the angular
Einstein radius, and so the proper motion:
\begin{equation}
\theta_\e = \frac{\theta_*}{\rho};\qquad \mu = \frac{\theta_\e}{t_\e}.
\label{e:theta_mu}
\end{equation}

If both $\theta_\e$ and $\pi_\e$ are measured, this immediately gives
both the lens mass, and the lens-source relative parallax \citep{gould92}
\begin{equation}
M = \frac{\theta_\e}{\kappa \pi_\e};\quad
\pi_{\rm rel} = \theta_\e\pi_\e, \qquad \kappa\equiv \frac{4G}{c^2\,\rm AU}
\end{equation}

Note that the ``proper motion'' in Equation (\ref{e:theta_mu}) is geocentric 
(i.e., in the instantaneous 
frame of the Earth at $t_{0,\rm par}$).  
To compare with heliocentric proper motions derived from astrometry at
successive epochs one must convert
\citep{janczak10}
\begin{equation}
\bmu_{\rm helio} -\bmu_{\rm geo} 
= \frac{{\bf v}_{\oplus,\perp}}{{\rm AU}} \pi_{\rm rel}
= \frac{{\bf v}_{\oplus,\perp}}{{\rm AU}} \frac{\theta_\e^2}{\kappa M}
\end{equation}
where 
${\bf v}_{\oplus,\perp}$ is the Earth velocity at $t_{0,\rm par}$
projected on the plane of the sky, and where 
$\bmu_{\rm geo} = \mu\bpi_\e/\pi_\e$.  Clearly this conversion can
only be carried out precisely if $\bpi_\e$ is known although the
rhs shows that the magnitude of the
difference can be strongly constrained even if there are only 
fairly crude limits on $\pi_\e$.

Note also, that future astrometric microlensing measurements may 
determine the (geocentric) vector proper motion $\bmu_{\rm geo}$ even 
if no parallax information is obtained 
\citep{hog95,miyamoto95,walker95}.

At this point we also introduce
\begin{equation}
t_*\equiv \rho t_\e,
\end{equation}
the source self-crossing time (in fact, a source radius crossing time).
Often, $t_*$ is used in place of $\rho$
as a fitting parameter.  
Note that sometimes $\rho$ is written as $\rho_*$
(in analogy to $t_*$), but for $\rho$ there is no need to subscript because
there are no competing quantities with this same name.  Hence, we advocate
simplifying the notation and dropping the subscript ``$*$''.

\subsection{Static Binary-Lens Parameters}

A static binary requires 3 additional parameters:
\begin{equation}
(q,s_0,\alpha_0) \qquad \rm (static\ binary\ parameters)
\end{equation}
These are the mass ratio of the two components ($q$=$m_2/m_1$=$M_2/M_1$),
their projected separation in units of the Einstein radius, 
and the
direction of lens-source relative motion (\ie, lens motion relative
to the source) with respect to the binary axis 
(which points from primary toward secondary). At the beginning one should 
specify which component of the binary one calls a ``primary''. 
(The angle $\alpha_0$ is counter-clockwise. The fractional mass
of the primary, $m_1$, is defined as $M_1/M$.)
There are many points to note.

First, the separation is frequently called ``$d$'' (rather than ``$s$'').
However, this is the standard symbol for the derivative operator,
and so should not be used for other quantities that are likely
to appear in the same expressions with this operator.

Second, heretofore, $s_0$ and $\alpha_0$ have been written simply as
$s$ and $\alpha$.  And indeed, for static binaries there is no
confusion in doing so, since these are time-invariant quantities.
And, for this reason, it remains appropriate to drop the ``0'' subscript
in static analysis.  Nevertheless, we introduce this subscript here
to maintain consistency with notation developed below.

Third, the choice of $t_{0,\rm par}$ is no longer obvious.
For high-magnification events, it might be taken as the approximate
time of closest approach to the center of magnification, which
closely parallels the choice for point lens events.  But it might
also be chosen to be a particularly well defined time, like a caustic
crossing.  In any case, since it is not even approximately obvious,
it must be specified.

Fourth, in generalizing from point lens to binary lens, there is
no longer a unique system ``center'' by which to define $u_0$ and
$t_0$.   However, since this does not present substantive problems
until there is orbital motion, we defer discussion of this point
until the next section.

Fifth, in the absence of parallax effects, static binaries are
subject to an exact degeneracy
\begin{equation}
(u_0,\alpha_0)\rightarrow -(u_0,\alpha_0)
\qquad \rm (binary\ degeneracy).
\end{equation}
Moreover, even for orbiting binaries, it is always 
possible to express solutions in a form with
$u_0>0$ and $0\leq \alpha_0 < 2\pi$.  Hence, in our view, negative
$u_0$ should be reserved for solutions that include parallax.

Finally, even if parallax is incorporated into the solution,
static binaries will be subject to a ``static binary ecliptic degeneracy''
\begin{equation}
(u_0,\alpha_0,\pi_{\e,\perp})\rightarrow -(u_0,\alpha_0,\pi_{\e,\perp})
\qquad \rm (static\ binary\ ecliptic\ degeneracy).
\end{equation}

Note that for lenses seen toward the Galactic Bulge, the directions
of positive ($\pi_{\e,\parallel},\pi_{\e,\perp}$) are typically
(West,North) for austral summer and autumn or (East,South) for
austral winter and spring.  Hence, from a practical standpoint,
it is often easy to locate degenerate $\pi_{\e,\perp}$ solutions by
seeding with a sign reversal of $\pi_{\e,N}$.

\subsection{Binary-Lens Parameters: Projected Orbital Velocity}
\label{a:orbit2d}

We begin by introducing the components of orbital motion
\begin{equation}
\bgamma\equiv (\gamma_\parallel,\gamma_\perp)
\qquad {\rm (at\ }t_{0,{\rm kep}}\rm ),
\end{equation}
which are the instantaneous components of velocity
of the secondary relative to the primary, respectively  parallel 
and perpendicular to the primary-secondary axis at a fiducial
time $t_{0,{\rm kep}}$.
These are essentially $\gamma_\parallel = (ds/dt)/s_0$ and 
$\gamma_\perp= - d\alpha/dt$, and they are detected effectively
from the changing shape and changing orientation of the caustic,
respectively.  In this form, their relation to the projected
physical orbital velocity is particularly simple,
\begin{equation}
\Delta{\bf v}= D_l\theta_\e s_0\bgamma,
\end{equation}
where $D_l$ is the distance to the lens.
Note that $(\gamma_\parallel,\gamma_\perp)$ is a right-handed system
on the plane of the sky, just like (N,E) and 
$(\pi_{\e,\parallel},\pi_{\e,\perp})$

The introduction of these two parameters brings with them
two degeneracies. First, allowing for orbital motion
reintroduces the $\pi_{\e,\perp}$ continuous degeneracy.
This appeared originally for point lenses because of
rotational symmetry, which is the physical reason that
$\pi_{\e,\perp}$ was fourth-order in time while $\pi_{\e,\parallel}$
was third-order (see Section \ref{s:pointlenshigh}).
Now it reappears because of a degeneracy
between $\pi_{\e,\perp}$ and the rotational degree of freedom
of orbital motion, i.e., as a correlation between
$\pi_{\e,\perp}$ and $\gamma_\perp$ (cf. \S\ref{s:piEN_omega}).  Second, to the degree
that the ecliptic degeneracy is present (i.e., to the degree that
the acceleration of the Earth does not change direction during the
event), it takes the form
\begin{equation}
(u_0,\alpha_0,\pi_{\e,\perp},\gamma_\perp)
\rightarrow -(u_0,\alpha_0,\pi_{\e,\perp},\gamma_\perp)
\qquad \rm (orbiting\ binary\ ecliptic\ degeneracy).
\label{e:orb_ecl_deg}
\end{equation}

Logically, $t_{0,{\rm par}}$ and $t_{0,{\rm kep}}$ can be different,
and so we have defined them separately.  There may be cases for
which one would want them to be different, but we cannot think of
any.  Therefore, we suggest in general that,
\begin{equation}
t_{0,\rm par}\equiv t_{0,\rm kep}, \qquad \rm (suggested).
\end{equation}
We will adopt this convention in what follows.
(This will also help to avoid confusion involving $\alpha_0$,
which is defined at $t_{0,{\rm par}}$, and $s_0$, which, together with other
phase-space parameters of the orbit, is defined at $t_{0,\kep}$.)

We now return to the problem of specifying a ``system center''.
For simplicity, let us begin by assuming that the
binary center of mass is chosen.  Then the problem of determining
$t_0$ and $u_0$ for a given model is identical to that of a single
lens (as discussed in Section \ref{s:pointlenshigh}) except that 
$t_{0,{\rm par}}$
may then differ substantially from $t_0$ 
(since it is convenient to choose $t_{0,{\rm par}}$ near the highest magnification), 
but this poses no
problem of principle. And, in particular, if $s<1$, then the
center of mass and center of magnification coincide, which greatly
simplifies the choices.  However, if $s>1$, one might for example
choose the system center to be the ``center of magnification''
and $t_{0,{\rm par}}$ to be approximately the time closest to this
center.  Then the difference between $t_{0,{\rm par}}$ and $t_0$ would
be small, but ``center of magnification'' would not in general
be in rectilinear motion (together with the center of mass).
In this example, the ``system center'' would be offset from the center of
mass by
\begin{equation}
\Delta(\xi,\eta)= \biggl[\frac{q}{1+q}\biggl(\frac{1}{s(t_{0,{\rm par}})} - s(t_{0,{\rm par}})\biggr),0\biggr],
\end{equation}
where $(\xi,\eta)$ are the coordinates on the lens plane parallel and
perpendicular to the primary-secondary axis.  Note that this offset
properly describes the true position of the ``center of magnification''
only at the one, chosen time ($t_{0,\rm par}$) and not at the fit time
$t_0$.  Again, for the case just specified, these two times would
be very close, so there is little practical impact.  However, in
other cases, particularly for resonant caustics of roughly equal-mass
binaries, $t_{0,\rm par}$ might be taken to be near a caustic crossing
that is very far from $t_0$.  Then specification of this time
becomes extremely important.

For binary lenses with higher-order parameter measurements, one
almost always measures $\theta_\e$.  And, because $\gamma_\perp$
is often degenerate with $\pi_{\e,\perp}$, it is usually
(but not always) inappropriate
to attempt to measure $\bgamma$ without also measuring $\bpi_\e$.
This means that when $\bgamma$ is measured one can usually
calculate the ratio of kinetic to potential projected energy
\citep{batista11},
\begin{equation}
\frac{E_{\perp,{\rm kin}}}{E_{\perp,{\rm pot}}}
= \frac{\kappa M_\odot\pi_\e(|\bgamma|{\rm yr})^2 s_0^3}{
8\pi^2\theta_\e(\pi_\e + \pi_s/\theta_\e)^3}.
\label{e:kin_pot}
\end{equation}
Evaluation of this ratio requires specification of one
additional parameter, the source parallax $\pi_s$.  This
is usually known, at least approximately, since most sources
are in the Galactic Bulge.

This ratio {\it must absolutely} be less than unity, if the
system is bound.  Moreover, typically it is expected to be in
the range 0.25 -- 0.6.  Hence, evaluating this ratio provides
a good plausibility check on the solution.  If $\theta_\e$
is measured in mas and $\bgamma$ is measured in yr$^{-1}$
(as we advocate), then the numerical coefficient in Equation (\ref{e:kin_pot})
is $8.14/(8\pi^2)= 0.103$.

\subsection{Binary-Lens Parameters: Full Kepler Solutions}
\label{a:kep}

\begin{figure}
\plotone{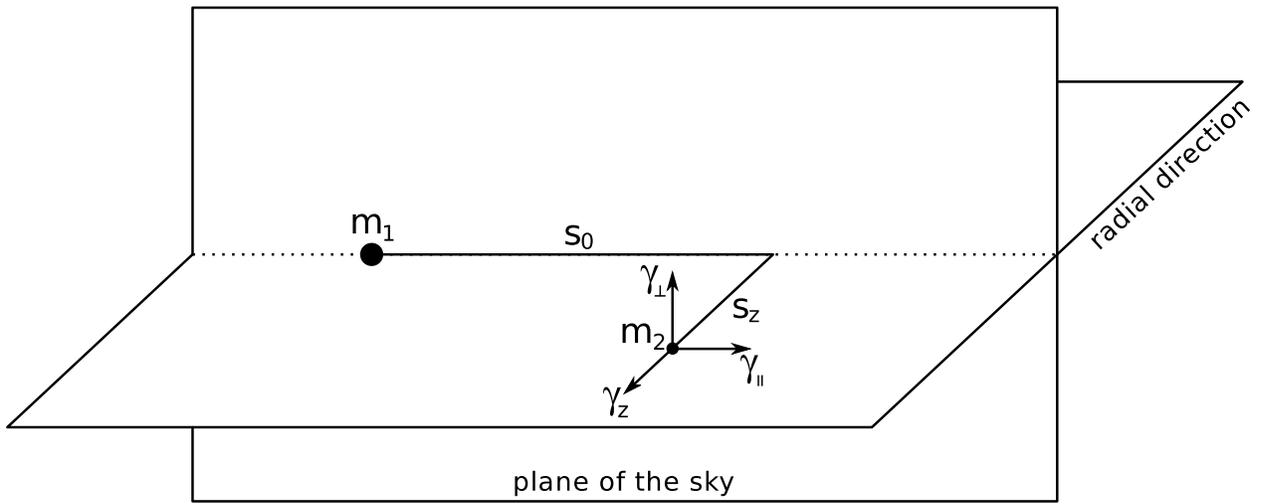}
\caption{Definition of the phase-space parameters 
$(s_0, 0, s_z, \gamma_\parallel, \gamma_\perp, \gamma_z)$
at $t_{0,\kep}$, describing motion of the secondary binary lens 
component ($m_2$) relative to the primary ($m_1$). The vertical plane 
is the plane of the sky.
\label{f:gamma}}
\end{figure}

Full Kepler solutions require two parameters in addition to
those already mentioned.  Since the parameters already specified
are in the Cartesian system, we advocate making the last two
parameters also Cartesian, namely the instantaneous position
and velocity in the direction perpendicular to the plane of the
sky at time $t_{0,\rm kep}$.  Specifically these are $s_z$ and $\gamma_z$,
which are defined so that the physical relative 3-dimensional 
position and velocity of the secondary relative to the primary are 
given by
\begin{equation}
\Delta {\bf r} = D_l\theta_\e (s_0,0,s_z), \qquad
\Delta {\bf v} = D_l\theta_\e s_0(\gamma_\parallel,\gamma_\perp,\gamma_z).
\end{equation}
(see Figure \ref{f:gamma}).
Recall that the $x$-axis is defined by the position of the secondary
relative to the primary at $t_{0,\rm kep}$.
These are right-handed triads, which means that the radial direction
points {\it toward the observer}, which is opposite to the convention
of radial-velocity (RV) work.  This conflict is not desirable but is
virtually unavoidable.  The identification of the $x$-axis with
the binary axis is extremely firmly rooted in microlensing tradition.
There is then only one way to maintain right-handed 2-dimensional
and 3-dimensional systems for $\bgamma$: the one adopted above.
If we were to choose a left-handed system, it would give rise
to substantial confusion in the calculation of orbits, which
involve several cross products.  Finally, it will be very rare
that any RV data are obtained on microlensing orbiting binaries.
When they are, one will just have to remember to reverse the sign.

Note that full Keplerian solution is possible for events with
$\pi_\e$ and $\theta_\e$ measured. For other events 
one can always use simplified 2-dimensional description 
of the orbital motion (\S\ref{a:orbit2d}), assuming that the 
fraction of the orbit travelled by the binary components 
during the event is small. 

There is one final degeneracy associated with two introduced
(radial) parameters.
As with visual binary orbits, there is a perfect degeneracy of
\begin{equation}
(s_z,\gamma_z) \rightarrow -(s_z,\gamma_z),
\end{equation}
which can only be broken with the aid of RV measurements.
Hence, microlensing solutions should choose {\it either}
$s_z\geq 0$ or $\gamma_z\geq 0$.  The choice will depend on
which quantity is more ``isolated'' from zero when both are
permitted free range.  However, this choice should definitely 
be made to avoid multiple representations of what are in fact
identical solutions.

\subsection{Summary}

Point-lens events are specified by up to 6 parameters
$(t_0,u_0,t_\e,\bpi_\e,\rho)$, where $\bpi_\e=(\pi_{\e,N},\pi_{\e,E})$
is the 2-dimensional parallax vector.   This is {\it always} a
geocentric system, {\it explicitly} if $\bpi_\e$ is specified and
{\it implicitly} if it is not.
Because $\bpi_\e$ is a geocentric quantity that depends on the
Earth's velocity, the definition of all of these quantities 
(except $\rho$) requires that a specific time $t_{0,\rm par}$ 
be adopted at which the Earth's velocity and position are evaluated.
The only exception is if $\bpi_\e$ is itself not specified.
These requirements carry over to binary lenses as well.

If $\rho$ is measured, then it is almost always possible to measure
$\theta_\e = \theta_*/\rho$ because $\theta_*$ can be determined
from the source flux parameters in two bands.

A total of eight parameters are required to specify a binary
orbit.  Two of these are secondary/primary mass ratio $q$ and the
total mass of the system $M=\theta_\e/\kappa\pi_\e$.  The
remaining 6 are the six Cartesian phase space coordinates
of the secondary relative to the primary.  In order to
relate microlensing parameters to physical units, one must
multiply by $r_\e\equiv D_l\theta_\e$.  The lens distance can
be expressed in terms of parameters already specified, plus the
source parallax: $D_l = {\rm AU}/(\pi_\e\theta_\e +\pi_s)$.
Then all but one of the six phase space coordinates appear
directly as microlensing parameters $(s_0,0,s_z)$ and 
$\bgamma=(\gamma_\parallel,\gamma_\perp,\gamma_z)$, for the
positions and velocities respectively, with all quantities
specified at $t_{0,\rm kep}$.  The missing degree of
freedom represented by the ``0'' in the spatial vector is
recovered in $\alpha_0$, which specifies the direction of lens-source
relative motion, measured with respect to the primary-secondary axis
at time $t_{0,\rm par}$.  We recommend $t_{0,\rm kep}= t_{0,\rm par}$.

For binary lenses, the ``system center'' must be given explicitly,
in order to define $t_0,u_0,t_\e$.  This center must be defined
as coordinates on the lens plane as determined at $t_{0,\rm par}$,
not at $t_0$.  See Section \ref{s:pointlensbasic}.

All coordinate systems are right-handed, either in two dimensions
[(N,E), $(\pi_{\e,\parallel},\pi_{\e,\perp})$, 
$(\gamma_\parallel,\gamma_\perp)$] or three dimensions
($\gamma_\parallel,\gamma_\perp,\gamma_z$).  This means, in
particular, that positive $\gamma_z$ points toward the observer.

All relative motion conventions are defined by the motion of the
lens (with the source thought of as fixed).  Thus, first, the
sign of $u_0$ is positive if the {\it lens} passes the source
on its right.  Second, $\phi_\pi={\rm atan2}(\pi_{\e,E},\pi_{\e,N})$
is the angle of {\it lens} motion, measured counter-clockwise
relative to North.  And third, $\alpha_0$ is the angle
of the {\it lens} motion, measured counter-clockwise relative
to the primary-secondary axis.

There are two approximate degeneracies, one discrete and one
continuous, which propagate through
the analysis as more parameters are added.  In their
``final form'' when binary orbital motion is modeled, the
discrete degeneracy is a generalized ``ecliptic degeneracy'',
which reverses the sign of the four parameters 
$(u_0, \pi_{\e,\perp}, \alpha_0, \gamma_\perp)$.  The continuous
degeneracy is between $\pi_{\e,\perp}$ and $\gamma_\perp$
(with $u_0$ maintaining the same sign).  These degeneracies
persist in simpler form when some of these parameters are
not measured.  Finally there is a perfect degeneracy that
reverses the signs of $(s_z,\gamma_z)$, so that the solutions
must be restricted to either $s_z\geq 0$ or $\gamma_z\geq 0$
to avoid duplicating identical solutions.

%

\section{Transformation between microlensing and Keplerian orbital parameters}
\label{a:jacobian}

In the process of model optimization we describe every solution
by a set of 15 `microlensing' parameters
($t_0$, $u_0$, $t_E$, $\rho$, $\bm{\pi}_\e$, $\theta_*$, 
$s_0$, $\alpha_0$, $s_z$, $\bm{\gamma}$, $\pi_s$, $q$).
In order to derive positions of the binary component at every given 
time we need to know the set of Keplerian parameters. They are also used
for introducing priors on the shape of the orbit since our intuition
works better in the space of orbital elements rather than cartesian
phase-space parameters. In this section we present the formulae we use
to transform microlensing parametrization into orbit elements.

As discussed by \citet{batista11} for the case of circular orbits,
the full Jacobian of the transformation from microlensing to
physical coordinates, factors into two Jacobians
\begin{equation}
\begin{split}
j_{\rm full}& =
\left\lVert\frac{\partial({\rm physical})}{\partial({\rm microlensing})}
\right\rVert =j_\kep\cdot j_{\rm gal}\\
j_\kep& =\left\lVert\frac{\partial(e, a, t_\peri, \Omega_\node, i, \omega_\peri)}
{\partial(s_0, \alpha_0, s_z, \gamma_\parallel, \gamma_\perp, \gamma_z)}\right\rVert \\
j_{\rm gal}& =\left\lVert\frac{\partial(M,D_l,\bmu)}
{\partial(\theta_\e,t_\e,\bpi_\e)}\right\rVert =
\frac{2\pi_{\rm rel} M\mu^2}{t_\e \theta_\e\pi_\e^2}\frac{D_l^2}{{\rm AU}}.
\end{split}
\label{e:jac}
\end{equation}
Here we focus on $j_\kep$.
(Note that, as remarked by \citet{batista11}, strictly speaking one should 
consider the parameters $\theta_*$ and $\rho$ separately, rather than
$\theta_\e=\theta_*/\rho$, but as these parameters barely vary over the
chain, this makes essentially no difference.). The priors
in the microlensing coordinates (used inside 
the MCMC routine in the transition probability, or used for weighting the 
likelihoods obtained from the chain) will be the priors in physical
parameters multiplied by the Jacobian ($j_{\rm full}$).

\subsection{Position parameters}
We have two systems of coordinates, the first one is
relative to the plane of the orbit and the second is related to
the microlensing event.
The microlensing system, as described in Appendix \ref{a:notation},
has its first axis set by the binary axis projected onto the
plane of the sky, and the third axis toward the observer. For
fitting, we use positions (${\bf s}$) and velocities ($\bm{\gamma}$) in 
microlensing units. However for the sake of this Appendix we will 
be using positions and velocities $({\bf r}, {\bf v})$ in physical units 
of $\rm{AU}$ and $\rm{AU}\, \rm{yr}^{-1}$. A simple conversion
between these units is given by ${\bf r} = {\bf s} \cdot D_l \theta_\e$ 
and ${\bf v} = r_x \bm{\gamma}$, where $D_l$ and $\theta_\e$ can be
calculated at any time from the set of microlensing parameters.

In the system of coordinates relative to the orbit, we set 
the first axis as the direction of periastron, 
and the third axis as that of the angular 
momentum vector. The position of the body (${\bf r}'$) in this system is 
given by:
\begin{align}
\br' & = \left(\begin{array}{c}
\cos \cE - e\\
\sqrt{1-e^2}\sin \cE\\
0
\end{array} \right) a
&
r' & = a(1-e \cos\cE)
\label{e:r}
\end{align}
where $a$ is the semi-major axis, $e$ is the eccentricity and $\cE$ is the 
eccentric anomaly, 
which is defined as an  implicit function of time $t$ 
by the Kepler equation, 
\begin{equation}
n (t-t_\peri) = \cE - e \sin \cE,
\qquad n\equiv \sqrt{\frac{GM}{a^3}}.
\label{e:kepler}
\end{equation}
Here $M$ is the system mass, $n$ is the mean motion, and
$t_\peri$ is the time of periastron.

The velocity in the plane of the orbit $({\bf v}')$ is given by
the derivative of position:
\begin{equation}
{\bf v}' = \frac{\dd{\bf r}'}{\dd t} = \left(\begin{array}{c}
- \sin \cE \\
 \sqrt{1-e^2} \cos \cE\\
0
\end{array} \right) a \frac{\dd\cE}{\dd t},
\qquad
\frac{\dd\cE}{\dd t} = \frac{n}{1-e\cos\cE},
\label{e:velocity}
\end{equation}
where the last result is found by implicit differentiation of
Equation~(\ref{e:kepler}).  Implicit differentiation also yields
\begin{equation}
\frac{\partial\cE}{\partial t_\peri} =
\frac{-n}{\sin\cE}\frac{\partial\cE}{\partial e} =
\frac{2 a}{3\Delta t}\frac{\partial\cE}{\partial e} = 
-\frac{\dd\cE}{\dd t}
\label{e:implicits}
\end{equation}

\subsection{Jacobian of the transformation}
The equations for ${\bf r}'$ and ${\bf v}'$ will be useful for calculating 
relative positions of the binary component at any time and for
evaluating the Jacobian of the transformation between the phase-space
coordinates used in the microlensing fit and the orbital elements 
$(j_\kep)$. This in turn will be used to construct the full Jacobian 
(\ref{e:jac})
used for weighting the solution, together with the priors on all `physical'
parameters (cf. \S\ref{s:orbital}):
\begin{multline}
j_\kep^{-1} =\left\lVert\frac{\partial(s_0, \alpha_0, s_z, \gamma_\parallel, \gamma_\perp, \gamma_z)}
{\partial(e, a, t_\peri, \Omega_\node, i, \omega_\peri)}\right\rVert =\\
(D_l \theta_\e)^{-6} s_0^{-4} \left\lVert\frac{\partial(r_x, r_y, r_z, v_x, v_y, v_z)}
{\partial(e, a, t_\peri, \Omega_\node, i, \omega_\peri)}\right\rVert 
= \frac{j_{{\rm ph}/\kep}}{(D_l \theta_\e)^6 s_0^4}
\end{multline}

\begin{figure}
\plotone{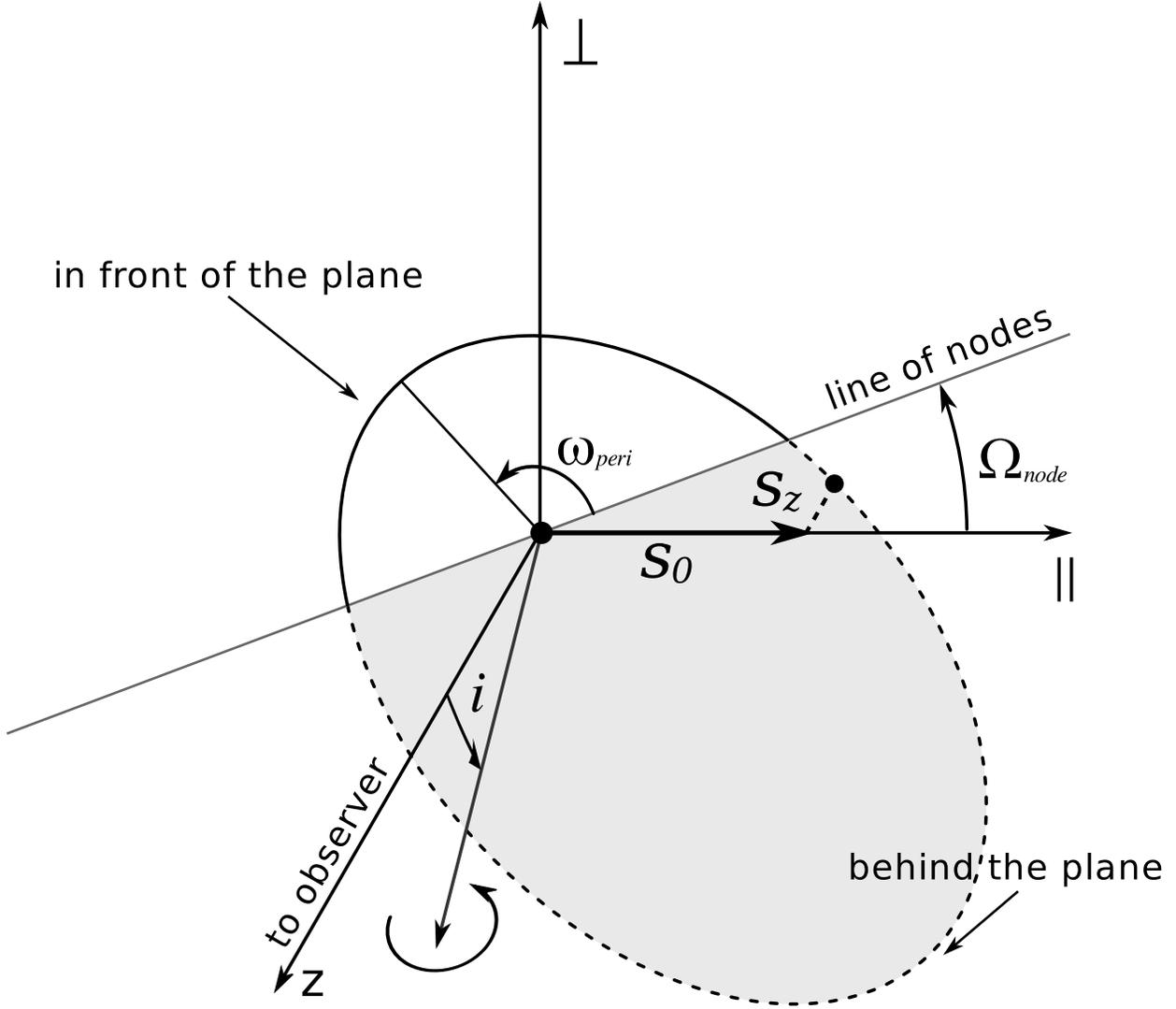}
\caption{The relative binary orbit which is rotated
by three Euler angles: longitude of nodes ($\Omega_{node}$), 
inclination ($i$), and argument of periapsis ($\omega_{peri}$). 
The binary components at the time $t_{0,\kep}$ are marked with 2 dots. 
The $z$ axis points toward the observer; axes marked with symbols
$\parallel$ and $\perp$ define the plane that is parallel to the 
plane of the sky and crosses the primary component of the binary.
The portions of the line that lie behind the plane are dashed.
The base coordinate system is
related to the microlensing event such that at the time 
$t_{0,\kep}$ the first axis coincides with the binary axis 
projected onto the plane of the sky. In the situation presented in this 
figure, the secondary is slightly behind the plane of the sky
$(s_z<0)$ and is about to reach ascending node; its velocity at 
$t_{0,\kep}$ has $\gamma_\parallel<0$, $\gamma_\perp>0$, $\gamma_z>0$.
\label{f:euler}}
\end{figure}

To calculate $j_{{\rm ph}/\kep}$ we need to evaluate derivatives of every 
component of $({\bf r}, {\bf v})$ with respect to every orbital element. 
To find $\partial {\bf r}/\partial (e,a,t_\peri)$ we will 
calculate $\partial {\bf r}'/\partial (e,a,t_\peri)$ in the plane
of the orbit and then rotate it to the microlensing system of 
coordinates using the rotation matrix $\Rot$ given by the three
orbital Euler angles ($\Omega_\node$ -- longitude of ascending node,
$i$ -- inclination, $\omega_\peri$ -- argument of periapsis):
\begin{equation}
{\bf r}=\Rot {\bf r}' ,\qquad
{\bf v}=\Rot {\bf v}'
\label{e:rotr}
\end{equation}
where:
\begin{equation}
\Rot = \Rot_z(\Omega_\node) \cdot \Rot_x(i) \cdot \Rot_z(\omega_\peri)
\label{e:r1r2r3}
\end{equation}
and $\Rot_x(\beta)$, $\Rot_z(\beta)$ are the matrix operators
of rotation by an angle $\beta$ around the first and third axes
respectively:
\begin{equation}
\Rot_x(\beta)=\left(\begin{array}{ccc}
 \cos \beta & -\sin \beta & 0 \\
 \sin \beta & \cos \beta & 0 \\
   0  &   0  & 1
\end{array}\right)
\quad
\Rot_z(\beta)=\left(\begin{array}{ccc}
 1 & 0 & 0\\
 0 &  \cos \beta & -\sin \beta \\
 0 &  \sin \beta & \cos \beta 
\end{array}\right)
\end{equation}
Figure \ref{f:euler} shows the orientation of the relative orbit 
with respect to the microlensing-coordinates system. Conventions 
used for Euler angles ($\Omega_\node, i, \omega_\peri$) can also 
be read from this Figure.

 From Equation~(\ref{e:implicits}) we obtain
\begin{equation}
\frac{\partial \br'}{\partial t_\peri} = - \bvv',
\qquad
\frac{\partial \bvv'}{\partial t_\peri} 
= - \ba' \equiv \frac{GM}{(r')^3}\br'
\end{equation}
\begin{equation}
\frac{\partial \br'}{\partial a} = -\frac{3\Delta t}{2a}\bvv' + \frac{\br'}{a},
\qquad
\frac{\partial \bvv'}{\partial a} = -\frac{3\Delta t}{2a}\ba' 
-\frac{1}{2} \frac{\bvv'}{a}
\label{e:tperi}
\end{equation}
\begin{equation}
\frac{\partial \br'}{\partial e} = \frac{\sin\cE}{n}\bvv' -
\left(\begin{matrix}1\cr \frac{e}{\sqrt{1-e^2}}\sin\cE \cr 0\end{matrix}\right)a,
\label{e:a}
\end{equation}
\begin{equation}
\frac{\partial \bvv'}{\partial e} = \frac{\sin\cE}{n}\ba'
+\frac{a\cos\cE}{r'}\bvv'
-\left(\begin{matrix}0\cr 1 \cr 0\end{matrix}\right)
\frac{a^2 ne\cos\cE}{r'\sqrt{1-e^2}},
\label{e:e}
\end{equation}
where $\ba'$ is the acceleration (not to be confused with $a$).

We then rotate these derivatives with $\Rot$:
\begin{equation}
\frac{\partial {\bf r}}{\partial (e,a,t_\peri)}=
\Rot \frac{\partial {\bf r}'}{\partial (e,a,t_\peri)},\quad
\frac{\partial {\bf v}}{\partial (e,a,t_\peri)}=
\Rot \frac{\partial {\bf v}'}{\partial (e,a,t_\peri)},
\end{equation}
and evaluate $\partial {\bf r}/\partial(\Omega_\node, i, \omega_\peri)$
\begin{align}
\frac{\partial {\bf r}}{\partial \Omega_\node} & = 
\frac{\partial \Rot}{\partial \Omega_\node} {\bf r}'  =
\frac{\partial \Rot_z(\Omega_\node)}{\partial \Omega_\node} \Rot_x(i) \Rot_z(\omega_\peri) {\bf r}' \\
\frac{\partial {\bf r}}{\partial i} & = 
\frac{\partial \Rot}{\partial i} {\bf r}'  =
\Rot_z(\Omega_\node) \frac{\partial\Rot_x(i)}{\partial i} \Rot_z(\omega_\peri) {\bf r}' \\
\frac{\partial {\bf r}}{\partial \omega_\peri} & = 
\frac{\partial \Rot}{\partial \omega_\peri} {\bf r}'  =
\Rot_z(\Omega_\node) \Rot_x(i) \frac{\partial\Rot_z(\omega_\peri)}{\partial \omega_\peri} {\bf r}',
\end{align}
and similarly for
$\partial {\bf v}/\partial(\Omega_\node, i, \omega_\peri)$.

From all derived derivatives we construct a $6 \times 6$ matrix  
and calculate its determinant, using
LUP matrix decomposition (LU decomposition with partial 
pivoting) on the lower and upper triangular matrices. Then, the 
determinant is given by simple multiplication of all diagonal 
elements of the triangular matrices. The Jacobian ($j_{{\rm ph}/\kep}$)
is given by the absolute value of the determinant.

It is useful to note that 
$j_{{\rm ph}/\kep}$ goes to zero proportional to $e \sin i$,
so for orbits close to circular or close to face-on, one must be 
careful about numerical problems when dividing by $j_{{\rm ph}/\kep}$.

\subsection{Deriving orbital parameters from phase-space parameters}
In this section we show how to translate phase-space parameters
in the microlensing system coordinates $({\bf r}, {\bf v})$ to orbital
elements $(e, a, t_\peri, \Omega_\node, i, \omega_\peri)$.

The specific orbital energy ($\varepsilon$) and specific relative
angular momentum (${\bf h}$) are conserved, so they
can be calculated given the 
values of ${\bf r}$ and ${\bf v}$ at any time.
\begin{align}
\varepsilon &= \frac{v^2}{2} - \frac{GM}{r} = -\frac{GM}{2a},
&
{\bf h}&={\bf r} \times {\bf v}.
\label{e:energy}
\end{align}
If we express ${\bf r}$ in the units of AU, ${\bf v}$ in the units 
of $\rm{AU}\, {yr}^{-1}$, and $M$ in $M_\Sun$ then $G \equiv 4\pi^2$.
The semi-major axis can be calculated from (\ref{e:energy}),
and the period can be derived from Kepler's Third Law:
\begin{align}
a &= - \frac{GM}{2 \varepsilon}
&
P &= 2\pi \sqrt{\frac{a^3}{GM}}.
\end{align}

Next, we find the unit vectors (versors) of the system of coordinates 
relative to the orbit, with $\vers{z}$ parallel to the angular momentum vector, 
and $\vers{x}$ parallel to the eccentricity vector\footnote{The
eccentricity vector is related to the Laplace-Runge-Lenz vector 
(${\bf A}$) by a scaling factor ${\bf e} = {\bf A}/(GMm)$}, ${\bf e}$, 
which is a constant of motion and points in the direction of periastron,
\begin{equation}
{\bf e} = \frac{{\bf v} \times {\bf h}}{GM} - \frac{{\bf r}}{r}.
\end{equation}
The eccentricity ($e=\lVert{\bf e}\rVert$) can also be calculated with
formula: $1-e^2 = h^2/(GMa)$.

The 3 versors are then given by,
\begin{equation}
\vers{x}=\frac{{\bf e}}{e},
\quad
\vers{z} = \frac{{\bf h}}{h},
\quad
\vers{y}=\vers{z}\times\vers{x}.
\end{equation}

From projections of the positional vector (${\bf r}$) onto the versors
($\vers{x}$ and $\vers{y}$) we calculate the true anomaly ($\nu$),
\begin{equation}
\sin\nu = \frac{\vers{y}\cdot{\bf r}}{r}
\quad
\cos\nu = \frac{\vers{x}\cdot{\bf r}}{r}
\end{equation}
then the eccentric anomaly,
\begin{equation}
\cos \cE = \frac{\cos \nu + e}{1+e \cos \nu},
\end{equation}
with the phase ambiguity in $\cE$ itself resolved by
\begin{equation}
0\leq\cE\leq\pi \Longleftrightarrow 0\leq\nu\leq\pi,
\qquad
-\pi<\cE<0 \Longleftrightarrow \pi <\nu<2\pi.
\end{equation}
This convention minimizes $|t_\peri - t_\kep|$.

The time of periastron can be derived from the inverted Equation
(\ref{e:kepler}) written for the current time ($t=t_{0,\kep}$):
\begin{equation}
t_\peri = t_{0,\kep} - \frac{\cE - e \sin \cE}{n}
\end{equation}

The coordinates of the 3 versors describing the orbital coordinate 
system ($\vers{x}$, $\vers{y}$, $\vers{z}$) give us the full
rotation matrix $\Rot$, defined earlier by Equations (\ref{e:r1r2r3}),
\begin{equation}
\Rot = \left(\begin{array}{ccc}
\vers{x} & \vers{y} & \vers{z}
\end{array} \right).
\end{equation}
We can use this matrix to find the angles $\Omega_\node$,
 $i$, and $\omega_\peri$ from the set of 9 Equations (\ref{e:r1r2r3})
for every element of $\Rot$. The inclination, which is always 
$0 \le i \le \pi$, is given by:
\begin{equation}
\cos i = \Rot_{33}.
\end{equation}
Then, the longitude of ascending node can be calculated from:
\begin{equation}
\cos \Omega_\node = -\frac{\Rot_{23}}{\sin i},
\quad
\sin \Omega_\node = \frac{\Rot_{13}}{\sin i},
\end{equation}
and the argument of periapsis is,
\begin{align}
\cos \omega_\peri &= \Rot_{11} \cos \Omega_\node + \Rot_{21} \sin \Omega_\node,
\\
\sin \omega_\peri &= (\Rot_{21} \cos \Omega_\node - \Rot_{11} \sin \Omega_\node)\cos i
                      + \Rot_{31} \sin i.
\end{align}
(For a special case of face-on orbits ($\sin i =0$) we assume 
$\omega_\peri \equiv 0$ and calculate $\Omega_\node$ from:
$\cos \Omega_\node = \Rot_{11}$ and $\sin \Omega_\node = \Rot_{21}$)

The relative position of the two binary components at any time, ${\bf r}(t)$,
can be calculated from Equation (\ref{e:rotr}) where $\Rot$ is 
a time-invariant matrix, and ${\bf r}'(t)$ is given by the 
Equations (\ref{e:r}) and  (\ref{e:kepler}).





\clearpage


\notetoeditor{(f:lc-all) in color}

\clearpage

\notetoeditor{(f:cmd) black and white for print, in color only in electronic version}

\clearpage

\notetoeditor{(f:chain-nat) in color}

\clearpage

\notetoeditor{(f:pien-thetadot-models) figure in color}

\clearpage

\notetoeditor{(f:pien-thetadot-chain) in black and white} 

\clearpage

\notetoeditor{(f:kepler) in color}

\clearpage

\notetoeditor{(f:gamma) in black and white}

\clearpage

\notetoeditor{(f:euler) in black and white}






\clearpage


\begin{deluxetable}{cccccc}
\tablecaption{Best-fit model parameters}
\tabletypesize{\scriptsize}
\tablehead{
\colhead{parameter}                & \colhead{unit}      & \colhead{parallax only}  &
\colhead{parallax + 2 par. motion} & \colhead{parallax + full orbit} & \colhead{parallax + full orbit} \\
\colhead{}                & \colhead{}      & \colhead{}  &
\colhead{}                & \colhead{}      & \colhead{(with priors)}
}
\startdata  
$\chi^2/dof$      &           &  370.29/361            &  352.7/359                        & 344.03/357 &   \\
fit parameters:   &           &                        &                                   &  &  \\
$t_0$             & $[HJD]$   &  $4917.266 \pm 0.008$  &  $4917.253_{-0.009}^{+0.011}$     & $4917.202_{-0.028}^{+0.059}$     &  $4917.252_{-0.009}^{+0.016}$   \\
$u_0/w$           &           &  $0.4295 \pm 0.0001$   &  $0.42921_{-0.00009}^{+0.00026}$  & $0.42963_{-0.00030}^{+0.00060}$  &  $0.42942 \pm 0.00030$   \\
$t_{\rm eff}$     & $[days]$  &  $4.517 \pm 0.011$     &  $4.72200 \pm 0.032$              & $4.92_{-0.12}^{+0.10}$           &  $4.708_{-0.032}^{+0.053}$ \\
$t_*$             & $[days]$  &  $0.1127 \pm 0.0003$   &  $0.11631 \pm 0.00080$            & $0.1182 \pm 0.0010$              &  $0.11595 \pm 0.00060$   \\
$\alpha$          & $[^o]$    &  $188.96 \pm 0.09$     &  $189.07 \pm 0.11$                & $189.59_{-0.26}^{+0.55}$         &  $189.08 \pm 0.14$   \\
$\pi_{\e,N}$      &           &  $0.45 \pm 0.03$       &  $-0.11_{-0.12}^{+0.10}$          & $-0.5 \pm 0.12$                  &  $-0.025 \pm 0.075$   \\
$\pi_{\e,E}$      &           &  $0.141 \pm 0.003$     &  $0.1468_{-0.0064}^{+0.0046}$     & $0.224_{-0.069}^{+0.019}$        &  $0.149 \pm 0.010$   \\
$\gamma_\perp$    & $[1/yr]$  &  ---                   &  $2.78_{-0.47}^{+0.41}$           & $4.2_{-0.5}^{+0.3}$              &  $2.3_{-0.3}^{+0.5}$  \\
$\gamma_z$        & $[1/yr]$  &  ---                   &  ---                              & $2.1 \pm 1.5$                    &  $1.7 \pm 0.6$ \\
$s_z$             & $[r_E]$   &  ---                   &  ---                              & $0.05_{-0.10}^{+0.17}$           &  $0.0 \pm 0.6$ \\
$\log q$          &           &  $-0.580 \pm 0.003$    &  $-0.5650 \pm 0.0025$             & $-0.5600 \pm 0.0060$             &  $-0.5650 \pm 0.0060$ \\
$\log w$          &           &  $-0.889 \pm 0.005$    &  $-0.8538 \pm 0.0044$             & $-0.8420_{-0.015}^{+0.007}$      &  $-0.8410 \pm 0.0080$ \\
$\gamma_\parallel$ & $[1/yr]$ &  ---                   &  $0.12_{-0.04}^{+0.08}$           & $0.34 \pm 0.30$                  &  $0.10 \pm 0.06$ \\
\hline
derived:          &           &                        &                                   &  &  \\
$s_0$             & $[r_E]$   &  $0.4149$              &  $0.4261$                         & $0.4299$                         & $0.4315$ \\
$q$               &           &  $0.263$               &  $0.272$                          & $0.275$                          & $0.272$  \\
$t_\e$            & $[days]$  &  $81.45$               &  $78.57$                          & $79.59$                          & $76.02$ \\
$u_0$             & $[r_E]$   &  $0.05546$             &  $0.06010$                        & $0.06182$                        & $0.06193$  
\enddata
\tablecomments{Best-fit parameters for 3 microlensing models 
with different treatment of lens orbital motion: 
1) with the lens as a static binary, 2) orbital motion of the lens 
projected onto sky is approximated by linear changes in time,
and 3) orbital motion of the lens is modeled using a full 
Keplerian orbit.
We give the values of the microlensing fit parameters
together with 1-$\sigma$ limits as obtained from MCMC.
The first three sets of parameters are derived from a MCMC run using only
likelihoods from the value of $\chi^2$ with no priors.
4) The solution including priors (on the orbital parameters and
on the properties of the lens) lies in 2.2 sigma
confidence limit derived from the light-curve $\chi^2$ only.
(All parameters represent positive $u_0$ solutions --
top region on Figure \ref{f:pien-thetadot-chain} -- 
which is slightly preferred. 
To obtain parameters for negative $u_0$ solutions use formula
(\ref{e:ecliptic}) from \S\ref{s:degeneracies})
}
\label{t:params}
\end{deluxetable}




\begin{thebibliography}{}
%
%
%
%
%


\bibitem[Alard(2000)]{alard00} Alard, C.\ 2000, \aaps, 144, 363 


\bibitem[Albrow et al.(2000)]{albrow00} Albrow, M.~D., et al.\ 
2000, \apj, 534, 894 

\bibitem[Albrow et al.(2009)]{albrow09} Albrow, M.~D., et al.\ 
2009, \mnras, 397, 2099 


\bibitem[Alcock et al.(2001)]{alcock01} Alcock, C., et al.\ 2001, 
\nat, 414, 617 


\bibitem[Afonso et al.(2000)]{afonso00} Afonso, C., et al.\ 2000, 
\apj, 532, 340 




\bibitem[An et al.(2002)]{an02} An, J.~H., et al.\ 2002, \apj, 572, 521 

\bibitem[An (2005)]{an05} An, J.~H.\ 2005, \mnras, 356, 1409 

\bibitem[An et al.(2007)]{an07} An, D., Terndrup, D.~M., 
Pinsonneault, M.~H., Paulson, D.~B., Hanson, R.~B., 
\& Stauffer, J.~R.\ 2007, \apj, 655, 233 


\bibitem[Batista et al.(2011)]{batista11} Batista, V., et al.\ 2011,
in prep.


\bibitem[Bennett et al.(2008)]{bennett08} Bennett, D.~P., et al.\ 2008,
\apj, 684, 663 

\bibitem[Bennett et al.(2010)]{bennett10} Bennett, D.~P., et al.\ 2010, 
\apj, 713, 837


\bibitem[Bensby et al.(2010)]{bensby10} Bensby, T., et al.\ 2010, 
\aap, 512, A41 


\bibitem[Berdyugina \& Savanov(1994)]{berdyugina94} Berdyugina, S.~V.,
 \& Savanov, I.~S.\ 1994, Astronomy Letters, 20, 755 


\bibitem[Bessell \& Brett(1988)]{bessell88} Bessell, M.~S., \& Brett, 
J.~M.\ 1988, \pasp, 100, 1134 


\bibitem[Bond et al.(2001)]{bond01} Bond, I.~A., et al.\ 2001, 
\mnras, 327, 868 



\bibitem[Chung et al.(2005)]{chung05} Chung, S.-J., et al.\ 2005, 
\apj, 630, 535 


\bibitem[Claret(2000)]{claret00} Claret, A.\ 2000, \aap, 363, 1081 



\bibitem[Dong et al.(2006)]{dong06} Dong, S., et al.\ 2006, 
\apj, 642, 842 

\bibitem[Dong et al.(2009a)]{dong09a} Dong, S., et al.\ 2009a, 
\apj, 695, 970 

\bibitem[Dong et al.(2009b)]{dong09b} Dong, S., et al.\ 2009b, 
\apj, 698, 1826 




\bibitem[Gaudi et al.(2008a)]{gaudi08a} Gaudi, B.~S., et al.\ 2008a, 
Science, 319, 927

\bibitem[Gaudi et al.(2008b)]{gaudi08b} Gaudi, B.~S., et al.\ 2008b, 
\apj, 677, 1268 


\bibitem[Gaudi(2010)]{gaudi10} Gaudi, B.~S.\ 2010, in Exoplanets, ed. S. Seager,
\url[http://arxiv.org/abs/1002.0332]{arXiv:1002.0332}

\bibitem[Gould(1992)]{gould92} Gould, A.\ 1992, \apj, 392, 442 

\bibitem[Gould et al.(1994)]{gould94} Gould, A., 
Miralda-Escude, J., \& Bahcall, J.~N.\ 1994, \apjl, 423, L105 

\bibitem[Gould(2000)]{gould00} Gould, A.\ 2000, \apj, 542, 785 

\bibitem[Gould(2004)]{gould04a} Gould, A.\ 2004, \apj, 606, 319 

\bibitem[Gould et al.(2004)]{gould04b} Gould, A., Bennett, D.~P., 
\& Alves, D.~R.\ 2004, \apj, 614, 404 

\bibitem[Gould(2008)]{gould08} Gould, A.\ 2008, \apj, 681, 1593

\bibitem[Gould(2009)]{gould09} Gould, A.\ 2009, Astronomical Society 
of the Pacific Conference Series, 403, 86 


\bibitem[Demarque et al.(2004)]{demarque04} Demarque, P., Woo, 
J.-H., Kim, Y.-C., \& Yi, S.~K.\ 2004, \apjs, 155, 667 



\bibitem[H\o{}g et al.(1995)]{hog95} H\o{}g, E., Novikov, I.~D., \& 
Polnarev, A.~G.\ 1995, \aap, 294, 287 


\bibitem[Houdashelt et al.(2000)]{houdashelt00} Houdashelt, M.~L., 
Bell, R.~A., \& Sweigart, A.~V.\ 2000, \aj, 119, 1448


\bibitem[Hwang et al.(2010)]{hwang10} Hwang, K.-H., 
et al.\ 2010, \apj, 723, 797 



\bibitem[Janczak et al.(2010)]{janczak10} Janczak, J., et al.\ 
2010, \apj, 711, 731 



\bibitem[Jaroszynski et al.(2005)]{jaroszynski05} Jaroszynski, 
M., et al.\ 2005, \actaa, 55, 159 


\bibitem[Jiang et al.(2004)]{jiang04} Jiang, G., et al.\ 2004, 
\apj, 617, 1307 



\bibitem[Kervella et al.(2004)]{kervella04} Kervella, P., Bersier, D., 
Mourard, D., Nardetto, N., Fouqu{\'e}, P., 
\& Coud{\'e} du Foresto, V.\ 2004, \aap, 428, 587 


\bibitem[Koz{\l}owski et al.(2007)]{kozlowski07} Koz{\l}owski, S., 
Wo{\'z}niak, P.~R., Mao, S., \& Wood, A.\ 2007, \apj, 671, 420 


\bibitem[Miyamoto \& Yoshii(1995)]{miyamoto95} Miyamoto, M., \& 
Yoshii, Y.\ 1995, \aj, 110, 1427 


\bibitem[Nishiyama et al.(2005)]{nishiyama05} Nishiyama, S., et al.
2005, \apjl, 621, L105 


\bibitem[Paczy{\'n}ski(1986)]{paczynski} Paczy{\'n}ski, B.\ 1986, \apj, 304, 1 



\bibitem[Park et al.(2004)]{park04} Park, B.-G., et al.\ 2004, \apj, 609, 166 



\bibitem[Pejcha \& Heyrovsk{\'y}(2009)]{2009ApJ...690.1772P} Pejcha, O., 
\& Heyrovsk{\'y}, D.\ 2009, \apj, 690, 1772 


\bibitem[Poindexter et al.(2005)]{poindexter05} Poindexter, S., 
Afonso, C., Bennett, D.~P., Glicenstein, J.-F., Gould, A., Szyma{\'n}ski, 
M.~K., \& Udalski, A.\ 2005, \apj, 633, 914 



\bibitem[Ram{\'{\i}}rez \& Mel{\'e}ndez(2005)]{ramirez05} 
Ram{\'{\i}}rez, I., \& Mel{\'e}ndez, J.\ 2005, \apj, 626, 446 




\bibitem[Ryu et al.(2010)]{ryu10} Ryu, Y.-H., 
et al.\ 2010, \apj, 723, 81


\bibitem[Schechter et al.(1993)]{schechter93} Schechter, P.~L., 
Mateo, M., \& Saha, A.\ 1993, \pasp, 105, 1342 



\bibitem[Smith et al.(2003)]{smith03} Smith, M.~C., Mao, S., 
\& Paczy{\'n}ski, B.\ 2003, \mnras, 339, 925 



\bibitem[Sumi et al.(2010)]{sumi10} Sumi, T., et al.\ 2010, \apj, 710, 1641 





\bibitem[Udalski et al.(2008)]{udalski08} Udalski, A., Szymanski, M.~K., 
Soszynski, I., \& Poleski, R.\ 2008, \actaa, 58, 69 


\bibitem[Walker(1995)]{walker95} Walker, M.~A.\ 1995, \apj, 453, 37 


\bibitem[Yelda et al.(2010)]{yelda10} Yelda, S., Ghez, A.~M., 
Lu, J.~R., Do, T., Clarkson, W., \& Matthews, K.\ 2010, 
\url[http://arxiv.org/abs/1002.1729]{arXiv:1002.1729}


\bibitem[Yoo et al.(2004)]{yoo04} Yoo, J., et al.\ 2004, 
\apj, 603, 139 


\bibitem[Wo{\'z}niak(2000)]{wozniak00} Wo{\'z}niak, P.~R.\ 2000, \actaa, 
50, 421 




\end{thebibliography}
\end{document}